\documentclass{article}

\usepackage{arxiv}

\usepackage[utf8]{inputenc} % allow utf-8 input
\usepackage[T1]{fontenc}    % use 8-bit T1 fonts
\usepackage{hyperref}       % hyperlinks
\usepackage{url}            % simple URL typesetting
\usepackage{booktabs}       % professional-quality tables
\usepackage{amsfonts}       % blackboard math symbols
\usepackage{nicefrac}       % compact symbols for 1/2, etc.
\usepackage{microtype}      % microtypography
\usepackage{lipsum}		% Can be removed after putting your text content
\usepackage{graphicx}
\usepackage{natbib}

\usepackage{amsmath} 
\usepackage{amsthm}
\usepackage{amssymb}
\usepackage{amsfonts}
\newcommand{\abs}[1]{\left|#1\right|}
\usepackage{float}
\usepackage{subcaption}
\usepackage{threeparttablex}
\usepackage{adjustbox}
\newtheorem{prop}{Proposition}
\newtheorem{assumption}{Assumption}
\newtheorem{definition}{Definition}
\newtheorem{Theorem}{Theorem}

\newcommand{\norm}[1]{\left\lVert#1\right\rVert}
\newtheorem{Lemma}{Lemma}
\DeclareRobustCommand{\varlambda}{\text{\usefont{OML}{txmi}{m}{it}\symbol{"15}}}
\usepackage{dsfont}
\newcommand{\fv}[2]{\num{#1}(\num{#2})} %formatted values

\usepackage{booktabs}      % For \toprule, \midrule, and \bottomrule
\usepackage{siunitx}       % For formatting numbers and table alignment
\usepackage{threeparttable} % For notes under the table
\usepackage{caption}       % To control captions
\usepackage{geometry}      % For setting margins

\title{A GARCH model with two volatility components and two driving factors}

\author{ Luca Vincenzo Ballestra\\
	Department of Statistical Sciences \\
	University of Bologna \\ Bologna, Italy \\
	\And
	  Enzo D'Innocenzo \\
	Department of Economics \\
	University of Bologna \\           Bologna, Italy \\
       \And
       Christian Tezza\thanks{
   Corresponding author. Dipartimento di Scienze Statistiche, Alma Mater Studiorum Università di Bologna, Via Belle Arti 41, 40126 Bologna, Italy, e-mail: \href{mailto:christian.tezza@unibo.it}{christian.tezza@unibo.it}.}\\
  Department of Statistical Sciences \\
	University of Bologna \\ Bologna, Italy \\
}

\renewcommand{\shorttitle}{\textit{arXiv} Template}

\hypersetup{
pdftitle={A template for the arxiv style},
pdfsubject={q-bio.NC, q-bio.QM},
pdfauthor={David S.~Hippocampus, Elias D.~Striatum},
pdfkeywords={First keyword, Second keyword, More},
}

\begin{document}
\maketitle

\begin{abstract}
	We introduce a novel GARCH model that integrates two sources of uncertainty to better capture the rich, multi-component dynamics often observed in the volatility of financial assets. This model provides a quasi closed-form representation of the characteristic function for future log-returns, from which semi-analytical formulas for option pricing can be derived. A theoretical analysis is conducted to establish sufficient conditions for strict stationarity and geometric ergodicity, while also obtaining the continuous-time diffusion limit of the model. Empirical evaluations, conducted both in-sample and out-of-sample using S\&P500 time series data, show that our model outperforms widely used single-factor models in predicting returns and option prices. The code for estimating the model, as well as for computing option prices, is made accessible in MATLAB language\footnote{The model code is available at the GitHub repository: \href{https://github.com/tezzachris/GARCH/tree/main/}{github.com/tezzachris/GARCH}}.
\end{abstract}

% keywords can be removed
\keywords{GARCH; Volatility component; Option pricing; Implied volatility; Two-factor model.}

\section{Introduction}

Empirical evidence indicates that the conditional volatility of financial assets is influenced by several stochastic components. Researchers have demonstrated the superiority of employing two factors over a single factor in capturing and modeling the rich and multifaceted dynamics of volatility. For example, \cite{alizadeh02}, \cite{bollerslev02}, and \cite{magistris15} support stochastic volatility specifications that incorporate two sources of uncertainty, highlighting the inadequacy of a simple one-factor volatility model to fully account for the dynamic dependencies observed in the daily volatility of exchange rates. Additionally, \cite{gallant99} and \cite{chernhov03} find that two-component volatility specifications provide higher accuracy than single-factor models when fitting equity market data and conclude that at least two components are necessary to capture the dynamics of volatility.

In addition, \cite{bates96}, \cite{taylor99}, \cite{fouque04}, \cite{schwartz09}, and \cite{fouque11} demonstrate that the introduction of a second volatility factor is crucial for capturing the behavior of the volatility surface implied by options. Moreover, \cite{bates00} and \cite{christoffersen09} propose two-factor stochastic volatility models that significantly outperform closely related one-factor models both in-sample and out-of-sample when applied to equity options data. Furthermore, \cite{christoffersen12} highlight the potential of multi-factor models for capturing time and cross-sectional variations in the implied volatility structure.

The aforementioned research documents the significantly superior performance achieved by multi-factor volatility models in continuous time. However, theoretical and empirical studies have also explored the multi-factor nature of volatility in discrete time. The generalized autoregressive conditional heteroskedasticity (GARCH) model, originally proposed by \cite{bollerslev86}, is one of the most popular and effective single-component models for filtering volatility in discrete time, as it inherently allows for direct estimation of volatility from historical returns. Consequently, given the complex and multifaceted nature of volatility observed empirically in financial markets, researchers have generalized the original GARCH model to incorporate two volatility components.

\cite{ding96} propose a novel approach to modeling volatility persistence by employing two GARCH components, demonstrating a significant likelihood improvement over the standard GARCH model when applied to S\&P500 daily returns data. Another two-component GARCH model is developed by \cite{engle99}, who provide evidence in support of volatility decomposition into two components by investigating the US and Japanese stock equity markets. \cite{adrian08} further advance this area by proposing a GARCH model with short-run and long-run components that significantly outperforms one-component GARCH-type models. Leveraging this approach and building on the work of \cite{engle99}, \cite{christoffersen08} introduce a Heston-Nandi GARCH model with two volatility components, demonstrating excellent option pricing performance compared to a one-component GARCH model. The model of \cite{christoffersen08} was generalized by \cite{christoffersen14} and \cite{bormetti15} by also including the realized variance. Additional evidence supporting the effectiveness of two-component models on equity returns data is presented in \cite{christoffersen10} and \cite{conrad20}.

All the two-component models mentioned so far have considered a single innovation term in the return equation. However, there is no inherent rationale to assume that both volatility components are driven by the same stochastic factor. In contrast, \cite{fouque00} and \cite{fouque03} extensively document the presence of two distinct volatility scales in returns driven by different random processes. Employing two different innovation terms seems appropriate, as the volatility components may relate to diverse sources of uncertainty. For example, \cite{adrian08}, \cite{engle08}, and \cite{engle13}, who propose two-component models with single innovation factors, indicate that one volatility component captures market skewness risk, which may be interpreted as a measure of the tightness of financial constraints, while the other volatility component models business cycle risk. 

This underscores the advantage of employing a different stochastic factor for each volatility component. Notably, \cite{zhang24} recently developed a non-affine model utilizing two variance components with two different stochastic factors by incorporating the realized variance. However, their approach implicitly assumes that both variance components are predictable, as both the returns and the realized variance are observable. Moreover, \cite{ghabani24} proposed a two-factor GARCH model that aims to capture the volatility of returns through two independent stochastic components. Nevertheless, the equations describing the two volatility components are independent, so the model does not account for spillovers between the volatility components, which 
may limit its ability to fully capture the complexities of the interaction mechanisms driving volatility dynamics. %Finally, theoretical matters regarding ergodicity and strict stationarity of the model are not analyzed.

Therefore, in the present work, filling a gap in the GARCH literature, we propose a GARCH model that incorporates two distinct components, each driven by an independent and unobservable innovation factor. In particular, by including two different stochastic factors, consistent with \cite{engle08}, \cite{adrian08}, and \cite{engle13}, we allow the model to effectively adapt to diverse market conditions and discern different market trends. For instance, as evidenced by the empirical analysis conducted in this paper, we can accurately identify two volatility components with different reactions to external shocks. Furthermore, our approach provides flexible modeling of volatility spillovers, incorporating the influence of each volatility component's lagged value on the other, thereby capturing dynamic interactions between the two volatility components.

    From a mathematical standpoint, the model we propose is a bivariate affine specification that extends the one-factor GARCH model of \citet{heston00}, allowing for a closed-form expression of the moment-generating function, which allows for quasi closed-form derivatives pricing. Notably, our proposed framework nests the recent model by \citet{ghabani24}, as it allows for a richer interaction structure between the volatility components. Additionally, the continuous-time limit of our volatility process recovers the popular stochastic volatility specification by \citet{christoffersen09}. Importantly, this study is the first in the literature to derive conditions for geometric ergodicity and strict stationarity in a GARCH model driven by two stochastic factors and two volatility components with coupled dynamics.

We test the empirical performance of the proposed two-factor GARCH model by comparing it against closely related models such as the one-factor model of \cite{heston00}, the component GARCH model of \cite{christoffersen08}, and the two-factor GARCH model of \cite{ghabani24}. We conduct extensive in-sample and out-of-sample exercises, focusing on the S\&P500 total return time series. The results show that our model exhibits superior performance compared to the benchmark models in explaining the cross-section of equity returns. Specifically, the introduction of a second innovation term significantly enhances the empirical fit compared to single-innovation benchmarks. Furthermore, the advantage of incorporating volatility spillovers leads to a marked improvement in predictive accuracy, both in-sample and out-of-sample, when applied to returns data.

Finally, after deriving the risk-neutralized version of the proposed model, we utilize it for pricing options written on the S\&P500 index. The empirical findings show that the inclusion of a second factor generally enhances the option pricing performance compared to single-innovation benchmark models across various moneyness and maturity levels.

The remainder of the paper is organized as follows. In Section \ref{sec_model}, we introduce the GARCH model with two innovations, establish sufficient conditions ensuring ensuring strict stationarity and geometric ergodicity, and derive a continuous-time limit. Section \ref{sec_estimation} discusses the parameter estimation procedure in detail. In Section \ref{sec_benchmarks}, we briefly review some popular (affine) GARCH models that we use as benchmarks. Section \ref{sec_empirical} presents the empirical performance of the proposed model and the benchmarks when applied to the S\&P500 total return time series. In Section \ref{sec_options}, we perform the risk-neutralization, derive the expression for the characteristic function, and in Section \ref{sec_option_empirics}, we test the models in pricing options written on the S\&P500. Finally, Section \ref{sec_conclusion} concludes. All mathematical proofs are gathered in the appendix.

\section{Modeling Returns with Two Factors}\label{sec_model}

In this section, we introduce a novel bivariate GARCH model with two innovation factors to capture the different components of the volatility of log-returns. This model will be labeled {GARCH with two factors} (\textit{GARCH-2F}).

Let $S_t$ denote the price of a risky asset at time $t \in \mathbb{Z}$ and let us consider the total log-return $R_t = \ln (\frac{S_t + D_t}{S_{t-1}})$, including the dividend $D_t$. We model the return process as follows:
 
\begin{equation} \label{eq:1}
    R_t = \mu_t + \sqrt{v_{1,t}} Z_{1,t} + \sqrt{v_{2,t}} Z_{2,t},
\end{equation}

\noindent where $v_{1,t}$ and $v_{2,t}$ denote the time-varying variance components. Moreover, we specify the conditional expected return as $\mu_t =  r + \lambda (v_{1,t} + v_{2,t})$, where $r$ represents the (constant) risk free rate and $\lambda$ is the (constant) risk premia parameter. Unlike traditional GARCH models and similarly to continuous time stochastic
volatility models, the specification in \eqref{eq:1} contains two sources of risk, $Z_{1,t}$ and $Z_{2,t}$, and we assume $Z_{1,t} \overset{\mathrm{i.i.d.}}{\sim} N(0,1)$, $Z_{2,t} \overset{\mathrm{i.i.d.}}{\sim} N(0,1)$ with $Z_{1,t} \perp Z_{2,t}$ for all $t \in \mathbb{Z}$. Note that the (conditional) total variance of returns is equal to $v_{1,t} + v_{2,t}$.

We model the time-varying variance components via the following bivariate system of equations:

\begin{equation} \label{eq:2}
    \begin{bmatrix}
    v_{1,t} \\
    v_{2,t} 
    \end{bmatrix} =
    \begin{bmatrix}
        \omega_1 \\
        \omega_2
    \end{bmatrix} + 
    \begin{bmatrix}
        \beta_{11} & \beta_{12} \\
        \beta_{21} & \beta_{22} 
    \end{bmatrix} 
    \begin{bmatrix}
    v_{1,{t-1}} \\
    v_{2,{t-1}} \\
    \end{bmatrix} + 
    \begin{bmatrix}
    \alpha_{11} & 
    \alpha_{12} \\
    \alpha_{21} & \alpha_{22} \\
    \end{bmatrix}
    \begin{bmatrix}
       \left(Z_{1,{t-1}}-\gamma_1 \sqrt{v_{1,{t-1}}} \right)^2 \\
       \left(Z_{2,{t-1}}-\gamma_2 \sqrt{v_{2,{t-1}}} \right)^2 
    \end{bmatrix}.
\end{equation}

 Equation \eqref{eq:2} specifies two volatility processes governed by Heston-Nandi type GARCH(1,1) dynamics, see \cite{heston00}, allowing spillovers between each variance component and innovation component. The specification includes only one lag in order to keep a parsimonious structure, but further lags can be added. Note that there is no spillover if $\beta_{12} = \beta_{21} = \alpha_{12} = \alpha_{21} = 0$. Moreover, if  $\omega_2 = \beta_{12} =  \beta_{22} = \beta_{21} = \alpha_{12} = \alpha_{22} = \alpha_{21} = 0$ we obtain a model with a single volatility component driven by the standard Heston-Nandi dynamic, see \cite{heston00}.

\begin{assumption} \label{assumption_1}
To ensure the positivity of the variances we assume that the parameters of model equations \eqref{eq:2} satisfy the following conditions:
\begin{align} \label{eq_assumptions}
    &\omega_i > 0, \quad  \beta_{ij} \geq 0, \quad \alpha_{ij} \geq 0, \quad i,j=1,2, \\ \notag
    &\alpha_{11} + \alpha_{12} > 0, \quad \alpha_{21} + \alpha_{22} > 0,
    \\ \notag
    &\beta_{11} + \beta_{12} > 0, \quad \beta_{21} + \beta_{22} > 0.
\end{align}

\end{assumption}

\subsection{Stationarity and Ergodicity}

In this section, we derive sufficient conditions under which the model given by \eqref{eq:1} and \eqref{eq:2} generates stationary and ergodic trajectories.

We prove the geometric ergodicity of $ \{R_t, \boldsymbol{v}_t\}$, where $\boldsymbol{v}_t = (v_{1,t}, v_{2,t})^T$,  based on the Markov chain stability theory of \cite{nummelin84} and \cite{tweedie93}. To this aim, let us define  $\mathcal{D} = [\omega_1,\infty) \times [\omega_2,\infty)$ and let us consider the following (non-linear) state representation of model \eqref{eq:1}-\eqref{eq:2}:

\begin{align}
R_t &= \mu_t + {G}(\boldsymbol{v_t}, \boldsymbol{Z}_{t}), \\
\label{eq_markov}
\boldsymbol{v}_t &= \boldsymbol{F} (\boldsymbol{v}_{t-1}, \boldsymbol{W}_{t} ),    
\end{align}

\noindent where 
$\boldsymbol{W}_{t} = \boldsymbol{Z}_{t-1}$ and ${G}: \mathcal{D} \times \mathbb{R}^2 \rightarrow \mathbb{R}$, and $\boldsymbol{F}: \mathcal{D} \times \mathbb{R}^2 \rightarrow \mathcal{D}$, are defined as follows:

\begin{equation*}
    {G} (\boldsymbol{v}_{t}, \boldsymbol{Z}_{t} ) =
    \begin{bmatrix}
        \sqrt{v_{1,t}} & \sqrt{v_{2,t}}
    \end{bmatrix}\begin{bmatrix}
        Z_{1,t} \\ Z_{2,t}
    \end{bmatrix},
\end{equation*}

\begin{equation*}
    \boldsymbol{F} (\boldsymbol{v}_{t-1}, \boldsymbol{W}_{t} ) = \boldsymbol{\omega} + \boldsymbol{\beta} \boldsymbol{v}_{t-1} + \boldsymbol{\alpha}  \begin{bmatrix}
       \left(W_{1,t}-\gamma_1 \sqrt{v_{1,t-1}} \right)^2 \\
       \left(W_{2,t}-\gamma_2 \sqrt{v_{2,t-1}} \right)^2 
    \end{bmatrix},
\end{equation*}

\noindent where $\boldsymbol{\omega} = \begin{bmatrix}
    \omega_1 & \omega_2
\end{bmatrix}^T$, $\boldsymbol{\beta} = \begin{bmatrix}
        \beta_{11} & \beta_{12} \\
        \beta_{21} & \beta_{22} 
    \end{bmatrix} $ and $ \boldsymbol{\alpha} = \begin{bmatrix}
    \alpha_{11} & 
    \alpha_{12} \\
    \alpha_{21} & \alpha_{22} \\
    \end{bmatrix}$.

This process is a homogeneous Markov chain with state space $(\mathcal{D}, \mathcal{B})$, where $\mathcal{B}$ is the Borel $\sigma$-algebra on $\mathcal{D}$. Following Chapter 7 of \cite{tweedie93}, we define inductively a sequence of functions $\boldsymbol{F}_t$ for $t=1,2,3,\ldots$, by $\boldsymbol{F}_{t+1}(\boldsymbol{x},\boldsymbol{W}_1,\ldots,\boldsymbol{W}_{t+1}) = \boldsymbol{F}(\boldsymbol{F}_t(\boldsymbol{x},\boldsymbol{W}_1,\ldots,\boldsymbol{W}_{t}),\boldsymbol{W}_{t+1})$, so that, for any initial condition $\boldsymbol{v}_0 = \boldsymbol{x}$ and $t \geq 1$, we can use equation \eqref{eq_markov} recursively to obtain $\boldsymbol{v}_t = \boldsymbol{F}_{t}(\boldsymbol{x},\boldsymbol{W}_1,...,\boldsymbol{W}_{t})$.  If we replace the random disturbances $\{ \boldsymbol{W}_t\}$ with a
deterministic control sequence, say $\{ \boldsymbol{w}_t\}$, we obtain the so-called deterministic control model associated to the non-linear state space model \eqref{eq_markov}, see Chapter 7 of \cite{tweedie93} for more details.

We use $\mu_{\text{Leb}}(\cdot)$ to denote the Lebesgue measure and $P^n\left(\boldsymbol{x}, A \right) = P \left( \boldsymbol{v}_n \in A \middle| \boldsymbol{v}_0 = \boldsymbol{x} \right)$, for  $\boldsymbol{x} \in \mathcal{D}$ and $A \in \mathcal{B}$, to denote the $n$–step transition probability measure of the Markov chain $\boldsymbol{v}_t$. For $n=0$ we have $P^0(\boldsymbol{x},A) = \mathrm{1}_A(\boldsymbol{x})$, that is the indicator of the set A: $1_A(\boldsymbol{x})=1$ if $\boldsymbol{x} \in A$, $1_A(\boldsymbol{x})=0$ if $\boldsymbol{x} \not\in A$. For $n = 1$ the notation $P (\boldsymbol{x}, A)$ is used. We denote with $\norm{\cdot}$ the $L_2$-norm of any vector and matrix and by $\rho(\boldsymbol{A})$ the spectral radius of any square matrix $\boldsymbol{A}$, \textit{i.e.}, $\rho(\boldsymbol{A}) = \max \{ |\varphi_i|: \varphi_i$ is an eigenvalue of $\boldsymbol{A}$\}.

For later purposes, we introduce the autoregressive matrix of the model \eqref{eq:2}:

    \begin{equation} \label{eq_auto}
        \boldsymbol{B} = \begin{bmatrix}
        \beta_{11} + \alpha_{11} \gamma^2_1 & \beta_{12} + \alpha_{12} \gamma^2_2 \\
        \beta_{21} + \alpha_{21} \gamma^2_1 & \beta_{22} + \alpha_{22} \gamma^2_2
    \end{bmatrix}. 
    \end{equation}

We state the definitions of the main concepts related to stationarity and ergodicity. 

\begin{definition}[\cite{tweedie93}]
A Markov process $\{\boldsymbol{v}_t\}$ is said geometrically ergodic if there exist a probability measure $\pi$ on $(\mathcal{D},\mathcal{B})$, a constant $0 < c < 1$, and a $\pi$-integrable nonnegative measurable function $Q$ such that:
\begin{equation*}
    \norm{ P^n(\boldsymbol{x},\cdot) - \pi(\cdot) }_{TV} \leq c^n Q(\boldsymbol{x}), \quad  n \geq 0, \quad \forall \boldsymbol{x} \in \mathcal{D},
\end{equation*}

where $\norm{\cdot}_{TV}$ denotes the total variation norm, that is $\norm{\mu}_{TV} = \sup_{ |g| \leq 1} |\mu(g)|$ for any signed  measure $\mu$ on $\mathcal{B}$.
\end{definition}

\begin{definition}[\cite{tweedie93}] \label{def_2}
    The Markov chain $\{ \boldsymbol{v}_t \}$ is $\psi$-irreducible if there exists a measure $\psi$ on $\mathcal{B}$ such that, for all $\boldsymbol{x} \in \mathcal{D}$, whenever $\psi(A)>0$, there exists some $k>0$, possibly depending on both $A$ and $\boldsymbol{x}$, such that $P^k(\boldsymbol{x},A) > 0$. 
\end{definition}

If $\{ \boldsymbol{v}_t \}$ is $\psi$-irreducible there exists a maximal irreducibility measure $M$ on $(\mathcal{D}, \mathcal{B})$, \textit{i.e.}, an irreducibility measure such that all other irreducibility measures are absolutely continuous with respect to $M$ (\cite{tweedie93}). Moreover, we set $\mathcal{B}^+ = \{ A \in \mathcal{B} : M(A) > 0 \}$.

\begin{definition}[\cite{nummelin84}] \label{def_aperiod}
    A sequence $(E_0,\ldots,E_{d-1})$ of $d$ non-empty disjoint sets in $\mathcal{B}$ is called a $d$-cycle if for all $i=0,\ldots, d-1$ and all $\boldsymbol{x} \in E_i$:
    \begin{equation*}
        P(\boldsymbol{x}, E^c_j) = 0, \quad  j= i+1 \mod(d).
    \end{equation*}
    If $d=1$, the chain is called aperiodic, otherwise is periodic.
\end{definition}

\begin{definition}[\cite{nummelin84}]

A set $C \in \mathcal{B}^+$ is called small if there are $n \geq 1$, $k > 0$ and a positive measure $\nu(\cdot)$ such that 

\begin{equation*}
    P^n(\boldsymbol{x}, \cdot) \geq k \nu(\cdot), \quad \forall \boldsymbol{x} \in C.  
\end{equation*}
\end{definition}

The following lemma appropriately defines the measure $\psi$ so that we achieve $\psi$-irreducibility and aperiodicity under mild-assumptions. 

\begin{Lemma} \label{lemma1} If Assumption \ref{assumption_1} holds, $\rho(\boldsymbol{B}) < 1$, $\rho(\boldsymbol{\beta}) < 1$ 
    and $\det ({\boldsymbol{\alpha}}) \neq 0$, the Markov chain $\{ \boldsymbol{v}_t\}$ is $\psi$-irreducible, where $\psi(\cdot) = \mu_{\text{Leb}}\left(\cdot \cap \Omega  \right) $ denotes the Lebesgue measure restricted to the set $\Omega$ defined in equation \eqref{eq_omegone} below. Moreover, $\{ \boldsymbol{v}_t\}$ is aperiodic.
\end{Lemma}

The proof of the Lemma, together with the other proofs, is reported in the appendix of the article. Here we note that in the proof of Lemma \ref{lemma1} we also derive an explicit form for the density of the transition probability associated to \eqref{eq:2}, see equation \eqref{eq_trans_density}.

Studying the small sets of an irreducible chain allows us to derive its long-run probabilistic behavior. Specifically, to prove geometric ergodicity, we first establish the following proposition and preliminary lemma. 

\begin{prop} \label{prop_small}
    If the assumptions of Lemma \ref{lemma1} hold and using the irreducibility measure $\psi$, the set $C= \{ \boldsymbol{x} \in \mathcal{D} : \norm{\boldsymbol{x}} \leq r \}$,  for any $r > \Bar{r}$ where $\Bar{r}$ is specified in equation \eqref{eq_rbarrato} below,  is a small set. 
\end{prop}

\begin{Lemma} \label{ineq_a}
    If Assumption \ref{assumption_1} holds, we have:
            \begin{align*}
    \norm{\boldsymbol{v}_t} & \leq 
    \norm{
    \boldsymbol{\omega} + \left\{ \boldsymbol{B} +  \boldsymbol{b}(\boldsymbol{W}_t) \right \} \boldsymbol{v}_{t-1} + \boldsymbol{\alpha }
    \boldsymbol{f}(\boldsymbol{W}_t) },
\end{align*}
\noindent where \begin{equation*}
        \boldsymbol{f}(\boldsymbol{W}_t) = \begin{bmatrix}
                \left(\abs{\gamma_1} + \sqrt{\gamma^2_1 + W^2_{1,t}} \right)^2  \\
                \left(\abs{\gamma_2} + \sqrt{\gamma^2_2 + W^2_{2,t}} \right)^2 
            \end{bmatrix} \quad \text{and} \quad 
            \boldsymbol{b}(\boldsymbol{W}_t) =
            \begin{bmatrix}
            \alpha_{11} W^2_{1,t} \mathds{1}_{(\gamma_1 W_{1,t} <0)} & \alpha_{12}  W^2_{2,t} \mathds{1}_{(\gamma_2 W_{2,t} <0)}\\
                \alpha_{21}  W^2_{1,t} \mathds{1}_{(\gamma_1 W_{1,t} <0)}& \alpha_{22}  W^2_{2,t} \mathds{1}_{(\gamma_2 W_{2,t} <0)}
            \end{bmatrix}.    
            \end{equation*}
\end{Lemma}

Then, given that $ \{ \boldsymbol{v}_t \} $ is irreducible and aperiodic, an appropriate small set exists and we can provide sufficient conditions for the geometric ergodicity of the process $\{R_t, \boldsymbol{v}_t\}$.

\begin{Theorem} \label{theorem}

If the assumptions of Lemma \ref{lemma1} hold and $\varphi = \mathbb{E} \left[\norm{ \boldsymbol{B} + \boldsymbol{b} (\boldsymbol{W}_t)}  \right] < 1$, then 
$\{ R_t, \boldsymbol{v}_t  \}$ is geometrically ergodic.
\end{Theorem}

We note that a sufficient ergodicity condition analogous to $\varphi < 1$ is obtained also in \cite{hafner09}.

\subsection{Continuous Time Limit}\label{sec_cont_limit}

In this section, we derive the continuous time limit of the model in equations \eqref{eq:1}-\eqref{eq:2}. We adopt the convergence scheme of \cite{nelson90} and also use the same notation.

\begin{prop} \label{prop_cont}
    If Assumption 1 holds, as the time interval shrinks, 
    the continuous-time diffusion limit of \eqref{eq:1}-\eqref{eq:2} is given by the following stochastic differential equations:

\begin{align} \label{eq_cont_limit}
    dR_t &=  (r + \lambda (v_{1,t} + v_{2,t})) dt + \sqrt{v_{1,t}} dW_{1,t} + \sqrt{v_{2,t}} dW_{2, t}, \\ \notag
    dv_{1,t} &= (\omega_1 + \alpha_{11} + \alpha_{12} - \theta_{11} v_{1,t} - \theta_{12} v_{2,t}) dt + 2 \alpha_{11} \gamma_1 \sqrt{v_{1,t}} dW_{3,t} + 2\alpha_{12}  \gamma_2\sqrt{ v_{2,t}} dW_{4,t}, 
    \\ \notag
    dv_{2,t} &= (\omega_2 + \alpha_{21} + \alpha_{22} - \theta_{21} v_{1,t} - \theta_{22} v_{2,t}) dt + 2 \alpha_{21} \gamma_1\sqrt{v_{1,t}} dW_{3,t} + 2 \alpha_{22} \gamma_2\sqrt{ v_{2,t} } dW_{4,t},
\end{align}

where the only non-zero correlations between Brownian motions are given by \begin{equation*}
    \mathbb{E} \left[dW_{1,t}dW_{3,t}\right]= - \text{sign}({\gamma_1})dt , \quad \mathbb{E} \left[dW_{2,t}dW_{4,t}\right]= - \text{sign}({\gamma_2})dt.
\end{equation*}
\end{prop}

We note that the set of equations \eqref{eq_cont_limit} is analogous to the bivariate
square-root stochastic volatility specification considered, for example, in \cite{christoffersen09}.

\subsection{Three Nested Models}

In this section, we present three parsimonious GARCH specifications nested within the \textit{GARCH-2F} model. This enables a comparison between the full \textit{GARCH-2F} model and its nested versions, where spillovers are either restricted or entirely absent.

\subsubsection{The \textit{GARCH-2F}$\boldsymbol{\beta}$ Model}

In this model we partially remove the spillover between the two volatilities in Equation \eqref{eq:2}, by imposing: 

\begin{equation*}
    \beta_{12} = \beta_{21} = 0.
\end{equation*}

\subsubsection{The \textit{GARCH-2F}$\boldsymbol{\alpha}$ Model}

In this model we remove the spillover between each volatility and the innovation driving the other. This amounts to imposing: 

\begin{equation*}
    \alpha_{12} = \alpha_{21} = 0.
\end{equation*}

\subsubsection{The \textit{GARCH-2F}$\boldsymbol{\alpha \beta}$ Model}

In this model we remove all the spillover components given by the $\alpha$ and $\beta$ coefficients. This amounts to setting: 

\begin{equation} \label{eq_jef_model}
      \beta_{12} = \beta_{21} = \alpha_{12} = \alpha_{21} = 0.
\end{equation}

By implementing these restrictions, we can completely eliminate the spillovers between the two volatility components, resulting in a more parsimonious variance process. Notably, under the specification in \eqref{eq_jef_model}, we recover the interesting two-factor GARCH model introduced by \cite{ghabani24}.

\section{Parameter Estimation and Filtering} \label{sec_estimation}

In this section, we illustrate a feasible and computationally efficient approach to estimate model \eqref{eq:1}-\eqref{eq:2}. Equation \eqref{eq:1} contains two sources of unobservable uncertainty, $Z_{1,t}$ and $Z_{2,t}$, which implies that we cannot estimate the model as done in traditional GARCH models, rather we need a filtering method. Following \cite{christoffersen12}, we note that a key property of a filter is that the filtered states are equal to their expected values conditional on the relevant information set. Hence, we define the filtered estimates of ${Z}_{1,t}$ and ${Z}_{2,t}$ conditional to the filtration $\mathcal{F}_t = \sigma(\{R_t, R_{t-1}, \ldots \})$ at time $t$ as

\begin{equation} \label{eq_zfiltrato}
    \widetilde{Z}_{i,t} = \mathbb{E}[ {Z}_{i,t} \mid  \mathcal{F}_t], \quad i=1,2.
\end{equation}

Since $Z_{i,t}$ is normally distributed, then also $R_t$ is normally distributed so that the conditional expectation in equation \eqref{eq_zfiltrato} can be computed using a well-known result on Normal distributions conditioning, see Chapter 2 of \cite{bda16}, that is:

\begin{equation} \label{eq:p}
    \widetilde{Z}_{i,t} = \frac{1}{\sqrt{\widetilde{v}_{i,t}}}\left( \frac{R_t  - \widetilde{\mu}_t}{\widetilde{v}_{1,t} + \widetilde{v}_{2,t}} \right), \quad i=1,2,
\end{equation}

\noindent where  $\widetilde{\mu}_t = r + \lambda (\widetilde{v}_{1,t} + \widetilde{v}_{2,t})$ and $\widetilde{v}_{1,t}$ and 
$\widetilde{v}_{2,t}$ denoting the filtered estimates of the conditional variances 
${v}_{1,t}$ and ${v}_{2,t}$ respectively, which can be computed using equation \eqref{eq:2}  

\begin{equation} \label{eq_fil_h}
    \begin{bmatrix}
    \widetilde{v}_{1,t} \\
    \widetilde{v}_{2,t} 
    \end{bmatrix} =
    \begin{bmatrix}
        \omega_1 \\
        \omega_2
    \end{bmatrix} + 
    \begin{bmatrix}
        \beta_{11} & \beta_{12} \\
        \beta_{21} & \beta_{22} 
    \end{bmatrix} 
    \begin{bmatrix}
    \widetilde{v}_{1,{t-1}} \\
    \widetilde{v}_{2,{t-1}} \\
    \end{bmatrix} + 
    \begin{bmatrix}
    \alpha_{11} & 
    \alpha_{12} \\
    \alpha_{21} & \alpha_{22} \\
    \end{bmatrix}
    \begin{bmatrix}
       \left(\widetilde{Z}_{1,{t-1}}-\gamma_1 \sqrt{\widetilde{v}_{1,{t-1}}} \right)^2 \\
       \left(\widetilde{Z}_{2,{t-1}}-\gamma_2 \sqrt{\widetilde{v}_{2,{t-1}}} \right)^2 
    \end{bmatrix}.
\end{equation}

The log-likelihood function can now be constructed as the product of the conditional distributions across the sample. Specifically, conditional on $\widetilde{v}_{1,0}$ and $\widetilde{v}_{2,0}$, the log-likelihood function of returns is given by

\begin{equation} \label{eq_likelihoodreturns}
    \ell_{\text{returns}}(\boldsymbol{\theta}) = -\frac{T}{2} \log{(2\pi (\widetilde{v}_{1,t} + \widetilde{v}_{2,t}))} - \frac{1}{2} \sum_{t=1}^T \frac{(R_t - \widetilde{\mu}_t)^2}{\widetilde{v}_{1,t} + \widetilde{v}_{2,t}},
\end{equation}

\noindent where $T$ denote the length of the daily log-returns series.

\section{Benchmarks Models} \label{sec_benchmarks}

Given the affine and two-component structure of model \eqref{eq:1}-\eqref{eq:2}, we will compare it with the following models: the GARCH model developed by \cite{heston00}, hereafter \textit{GARCH-HN}, the two-component GARCH model introduced by \cite{christoffersen08}, hereafter \textit{GARCH-CJOW} model, and the \textit{GARCH-2F$\alpha\beta$} of \cite{ghabani24}. Both the \textit{GARCH-HN} and \textit{GARCH-CJOW} models employ a single innovation factor, and, for the reader's convenience, they are briefly recalled below.

\subsection{The \textit{GARCH-HN} Model}

The return and the volatility processes are defined as follows

\begin{align*}
    R_t &= r + \lambda v_{1,t} + \sqrt{v_{1,t}} Z_{t} , \\ 
    v_{1,t+1} &= \omega_1 + \beta_{11} v_{1,t} + \alpha_{11} (Z_{t}-\gamma_1 \sqrt{v_{1,t}})^2,
\end{align*}

\noindent where $Z_t  \overset{\mathrm{iid}}{\sim} N(0,1)$. 

\subsection{The \textit{GARCH-CJOW} Model}

The component GARCH model proposed by \cite{christoffersen08} comprises three equations, one for the return process, one for the long-term variance component $q_{t}$, and one for the short-term variance component $s_{t}$:

\begin{align*}
    R_t &= r + \lambda (q_{t}+s_t) + \sqrt{q_{t}+s_t} Z_{t}, \\
    q_{t+1} &= \omega_1 + \beta_{11} q_{t} + \alpha_{11} (Z^2_{t}-1-2\gamma_1 Z_{t} \sqrt{q_{t}+s_t}), \\
    s_{t+1} &= (\beta_{22}+\alpha_{22}\gamma^2_2)s_{t} + \alpha_{22} (Z^2_{t}-1-2\gamma_2 Z_{t}\sqrt{q_{t}+s_t}),
\end{align*}

\noindent where $Z_t \overset{\mathrm{iid}}{\sim} N(0,1)$.

\section{Empirical Results for Returns} \label{sec_empirical}

\begin{figure}[h]
    \centering
    \includegraphics[scale=0.6]{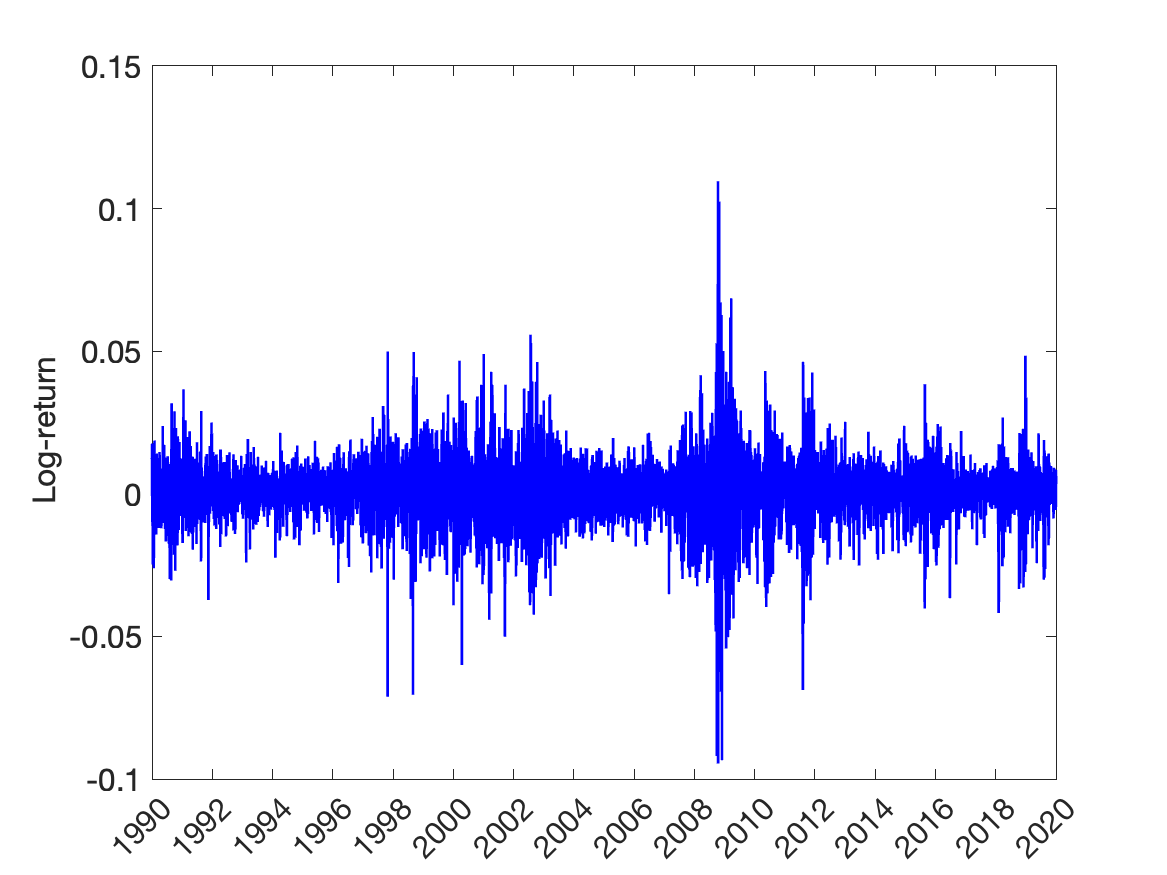}
    \caption{Daily time series of total S\&P500 log-returns from January 5, 1988 to December 29, 2023.}
    \label{fig:data}
\end{figure}

In this section, we present empirical results for the total S\&P500 daily adjusted log-returns time series. Our data span from January 5, 1988 to December 29, 2023 (9069 daily observations), as shown in Figure \ref{fig:data}.

\subsection{In-Sample Analysis}

The estimation procedure described in Section \ref{sec_estimation} requires, besides the time series of log-returns, also the risk-free interest rate, which we proxy using the 3-month US Treasury Bill rate. The data for the daily levels of the Adjusted Close Price for the S\&P500 total Returns series is retrieved from Refinitiv Datastream, whereas the data for the 3-month Treasury Bill rate are gathered from the Federal Funds Effective Rate (FRED) dataset. 

The in-sample results, based on the maximum likelihood estimation described in Section \ref{sec_estimation}, are illustrated in Table \ref{tab_insample}. To assess the model performance we consider the Akaike (AIC) and the Bayesian information criteria (BIC). To further assess the goodness-of-fit of our model specification we performed a likelihood ratio (LR) test for the $\textit{GARCH-2F}$ model and its three nested models versus the one factor \textit{GARCH-HN} model. The LR statistics are reported at the bottom of Table \ref{tab_insample}. 

As we may see, the $\textit{GARCH-2F}$ model and its nested versions provide superior goodness-of-fit to the returns data compared to the one-factor benchmark models. The LR statistic is significantly higher, and the AIC and BIC values are lower by at least 60 points. Moreover, the most parsimonious model, the $\textit{GARCH-2F}\alpha \beta$ proposed by \cite{ghabani24}, achieves 30 likelihood extra points compared to $\textit{GARCH-CJOW}$ and 130 points compared to $\textit{GARCH-HN}$. However, in comparison to $\textit{GARCH-2F}\alpha \beta$, results also highlight the significance of including volatility spillovers, either through $\alpha$ for the \textit{GARCH-2F$\beta$} model or the $\beta$ coefficient for the \textit{GARCH-2F$\alpha$} model specification.

We further assessed the volatility persistence for all the considered models. The persistences of the two variance components $v_{1,t}$ and $v_{2,t}$ in model \eqref{eq:1}-\eqref{eq:2} are defined as the two eigenvalues of the autoregressive matrix $\boldsymbol{B}$ in  \eqref{eq_auto}. Whereas, the persistences, of the \textit{GARCH-CJOW} model, as discussed in \cite{christoffersen08}, are equal to $\beta_{11} + \alpha_{11} \gamma^2_1$ for the one volatility component and $\beta_{22} + \alpha_{22} \gamma^2_2$ for the other volatility component. At the bottom of Table \ref{tab_insample} we present the persistences of the volatilities for all the models. A feature that characterize all the component models is that the component displaying the highest $\beta$ coefficient is also the one possessing the lowest reaction to innovation shocks, as indicated by the $\alpha$ coefficient. However, the persistence values do not differ significantly among the nested model specifications and are aligned to the peristences of the \textit{GARCH-CJOW} model.

%We note that a feature found also in \cite{alizadeh02} is that the estimated one-factor volatility persistence parameter is an average of the estimated persistence parameters from the two-factor model.

When analyzing the model parameters, we observe that for all considered models, the first volatility component returns a $\beta_{11}$ estimate around 0.9, which is commonly encountered in the GARCH literature (see \cite{heston00}, \cite{christoffersen08}). A similar pattern holds for $\alpha_{11}$ and $\gamma_1$. The parameter $\omega_1$ is largely non-significant across most models, except for the \textit{GARCH-CJOW}. In the component models, the second volatility component exhibits higher sensitivity to innovations, as reflected by $\alpha_{22}$ and $\gamma_2$. The autoregressive parameter $\beta_{22}$ for the second component is generally lower than $\beta_{11}$ and is not always significant in the \textit{GARCH-2F}, $\textit{GARCH-2F}\beta$ and $\textit{GARCH-2F}\alpha$ models.

If we consider the spillover effects between the two volatility components in the $\textit{GARCH-2F}$ and the $\textit{GARCH-2F}\beta$ models, the coefficients $\alpha_{12}$ and $\alpha_{21}$ are highly significant. In the $\textit{GARCH-2F}\alpha$ model, the spillovers through the $\beta_{12}$ and $\beta_{21}$ coefficients are also significant. Consequently, the more parsimonious \textit{GARCH-2F}$\alpha \beta$ specification proposed by \cite{ghabani24}, which eliminates spillovers between the two volatility components, is not fully capable of capturing the dynamics of return data.

\begin{table}[htbp]
\centering
\begin{threeparttable}
    \caption{Maximum likelihood estimation results.}
    \label{tab_insample}
    \begin{tabular}{r 
                    S[table-format=-3.3] 
                    S[table-format=-3.3] 
                    S[table-format=-3.3] 
                    S[table-format=-3.3] 
                    S[table-format=-3.3] 
                    S[table-format=-3.3] 
                    }
    \toprule
        &   {\textit{GARCH-HN}}           &   {\textit{GARCH-CJOW}}     & {\textit{GARCH-2F}}     &  {\textit{GARCH-2F$\beta$}}   &  {\textit{GARCH-2F$\alpha$}} &  {\textit{GARCH-2F$\alpha \beta$}} \\
    \midrule
    $\omega_1$           & 0 &  {1.115e-06}\tnote{***} & 0                & 0                      &   0                   & 0 \\
                        & {(8.849e-09)}             & {(2.193e-07)}     & {(6.925e-08)}            &  {(8.911e-09)}         & {(1.695e-08)} & {(2.792e-10)}\\
    \addlinespace
    $\omega_2$           &  &   & 0                & 0                      &   0                   & 0 \\
                        &             &    & {(5.268e-08)}            &  {(3.351e-09)}         & {(2.644e-08)} & {(6.743e-10)}\\
    \addlinespace
    $\alpha_{11}$       & {5.055e-06}\tnote{***} & {1.764e-06}\tnote{***}          & {2.967e-07}\tnote{**}      &  {2.966e-07}\tnote{***}           & {4.676e-07}\tnote{***}                  & {1.503e-07}\tnote{***}\\
                        &  {(9.089e-07)}           &  {(3.426e-07)}               & {(1.327e-07)}     & {(1.069e-08)}           & {(1.108e-07)}    & {(5.654e-10)}\\
    \addlinespace
    $\gamma_{1}$        &   169.418\tnote{***}     &   115.328\tnote{***}     & 107.640\tnote{***}          &  107.638\tnote{***}       & 145.708\tnote{**}      & 358.980\tnote{***}\\
                        &  {(17.864)}                &  {(26.286)}                & {(32.944)}       &         {(25.913)}                 & {(88.501)}
                        & {(6.042)} \\
    \addlinespace
    $\beta_{11}$        &   0.812\tnote{***}        &   0.991\tnote{***}         &  0.986\tnote{***}   & 0.986\tnote{***}         & 0.975\tnote{***}     & 0.978\tnote{***}\\
                        &  {(0.016)}                 &  {(0.001)}                    &  {(0.004)}            & {(0.004)}                  & {(0.014)}    & {(0.001)} \\
    \addlinespace
    $\alpha_{22}$       &                          &  {2.441e-06}\tnote{***}     & {5.334e-06} \tnote{**}     &  {5.333e-06}\tnote{***}  & {4.195e-06}\tnote{***} & {1.039e-05} \tnote{***}\\
                        &                          & {(7.818e-07)}               & {(8.491e-07)}    &    {(8.110e-07)}           & {(1.556e-06)}   
                        & {(2.069e-07)}\\
    \addlinespace
    $\gamma_{2}$        &                          &  369.261\tnote{***}         & 419.987\tnote{***}      &   420.007\tnote{***}      & 469.691\tnote{***}     & 154.782 \tnote{***}\\
                        &                          &  {(115.072)}                   & {(37.253)}               &  {(33.263)}                & {(85.395)}        & {(2.691)} \\
    \addlinespace
    $\beta_{22}$        &                          &  0.539\tnote{***}           &  {2.599e-07}             &  {1.607e-06}         & 0.011    & 0.684\tnote{***}\\
                        &                          & {(0.127)}                     &  {(8.989e-07)}           & {(1.903e-06)}                 & {(0.019)} 
                        & {(0.013)} \\
    \addlinespace
    $\alpha_{12}$       &                          &                             &  {2.921e-08} \tnote{***}        &  {2.921e-08}\tnote{***}   &           \\
                        &                          &                             &  {(1.155e-08)}                  & {(1.185e-08)}           &    \\
    \addlinespace
    $\alpha_{21}$       &                          &                             &  {1.716e-06}\tnote{***}                 & {1.716e-06}\tnote{***}  &            \\
                        &                          &                             &  {(6.431e-07)}               & {(6.072e-07)}           &     \\
    \addlinespace
    $\beta_{12}$        &                          &                             &  0           &                         & 0.005\tnote{***}               \\
                        &                          &                             &  {(2.107e-08)}       &                        &  {(0.001)}  \\
    \addlinespace
    $\beta_{21}$        &                          &                             &  0 &                         & 0.058\tnote{**}        \\
                        &                          &                             &  {(1.373e-08)}     &                  & {(0.030)} \\
    \addlinespace
    $\lambda$          &  1.101                   &  1.115\tnote{***}                      &  1.630               & 1.632        & 2.170      & 0.903  \\
                        & {(1.073)}                  & {(0.349)}                     & {(4.790)}               &  {(1.575)}                & {(1.830)}
                        & {(0.953)}\\
    \midrule
    Log-likelihood      & {29,667}     &     {29,761}  &   {29,831}                       & {29,831}    &    {29,812}  & {29,797}     \\
    AIC &  {-59,324} & {-59,506} & {-59,636} & {-59,640} & {-59,602} & {-59,576} \\ 
    BIC  &  {-59,288} & {-59,449} & {-59,543} & {-59,561} & {-59,523} & {-59,511} \\
    LR & 0 & {-} & 328 & 328 & 290 &  260 \\
    Persistence   & 0.96   &  {0.99 $\vert$ 0.87}  & {0.99 $\vert$ 0.94}  &  {0.99 $\vert$ 0.94}    &  {0.99 $\vert$ 0.93}  & {0.99 $\vert$ 0.93} \\
    \bottomrule
    \end{tabular}
    
    \begin{tablenotes}
        \item The estimation period spans from January 5, 1988 to December 29, 2023 (9069 observations). The standard errors, reported in parentheses, are computed by inverting the negative Hessian matrix evaluated at the optimum parameter values. [*],[**],[***]  denote statistical significance at the 0.1, 0.05, 0.01 levels, respectively.
    \end{tablenotes}
\end{threeparttable}
\end{table}

\begin{figure}[htbp]
\centering
\includegraphics[width=.333\textwidth]{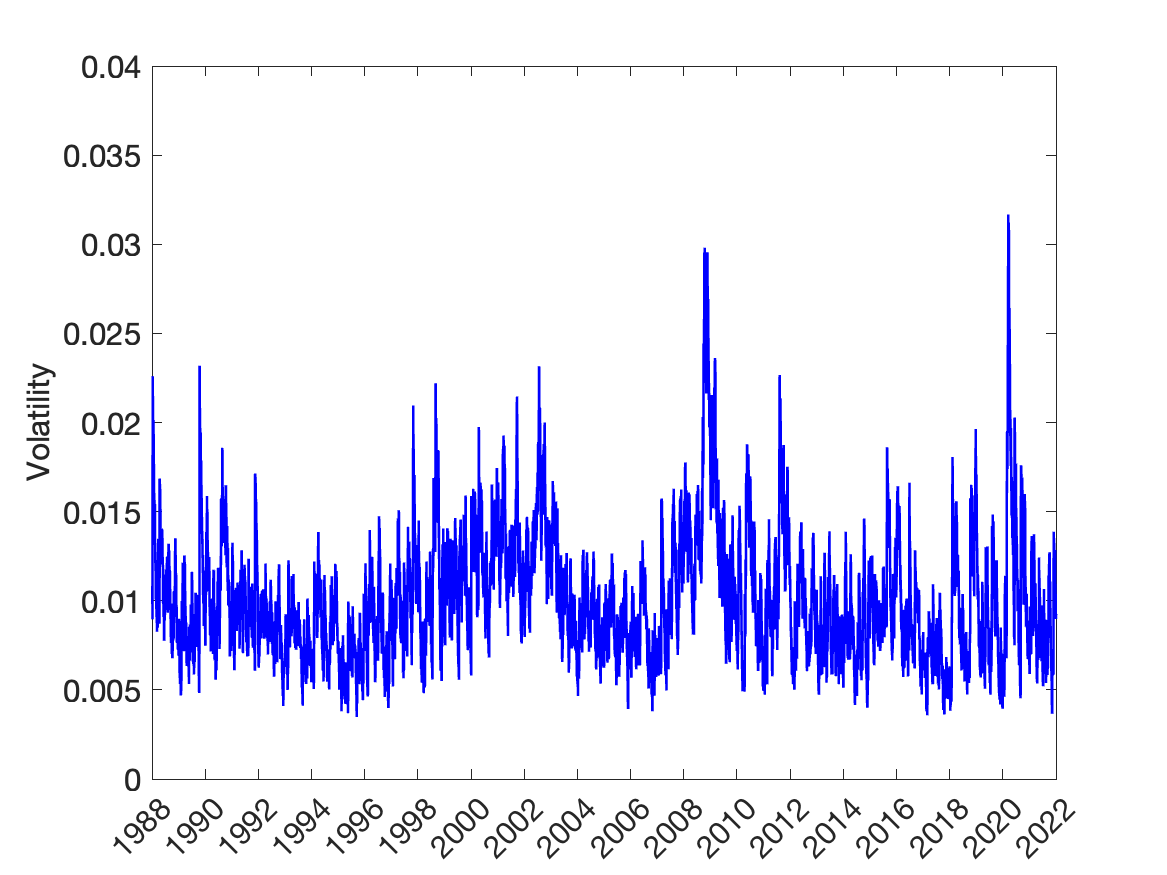}\hfill
\includegraphics[width=.333\textwidth]{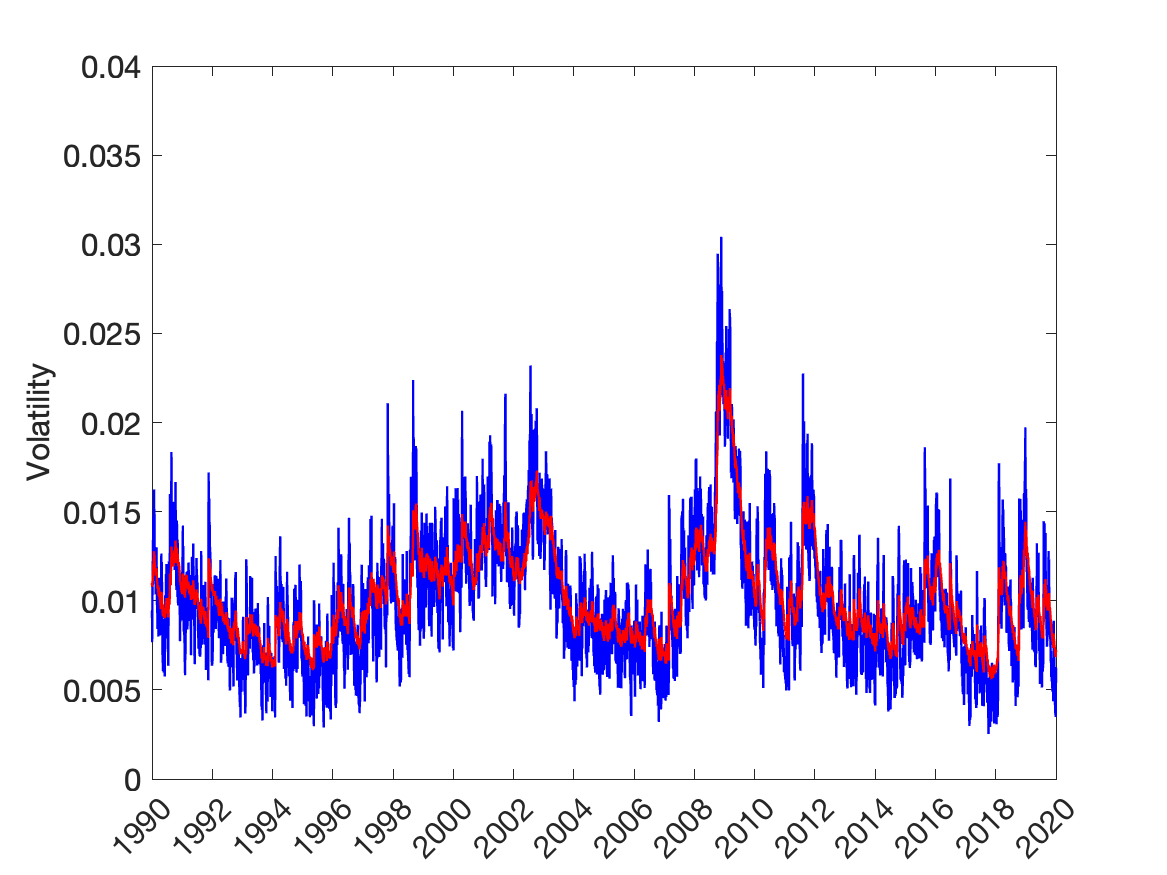}\hfill
\includegraphics[width=.333\textwidth]{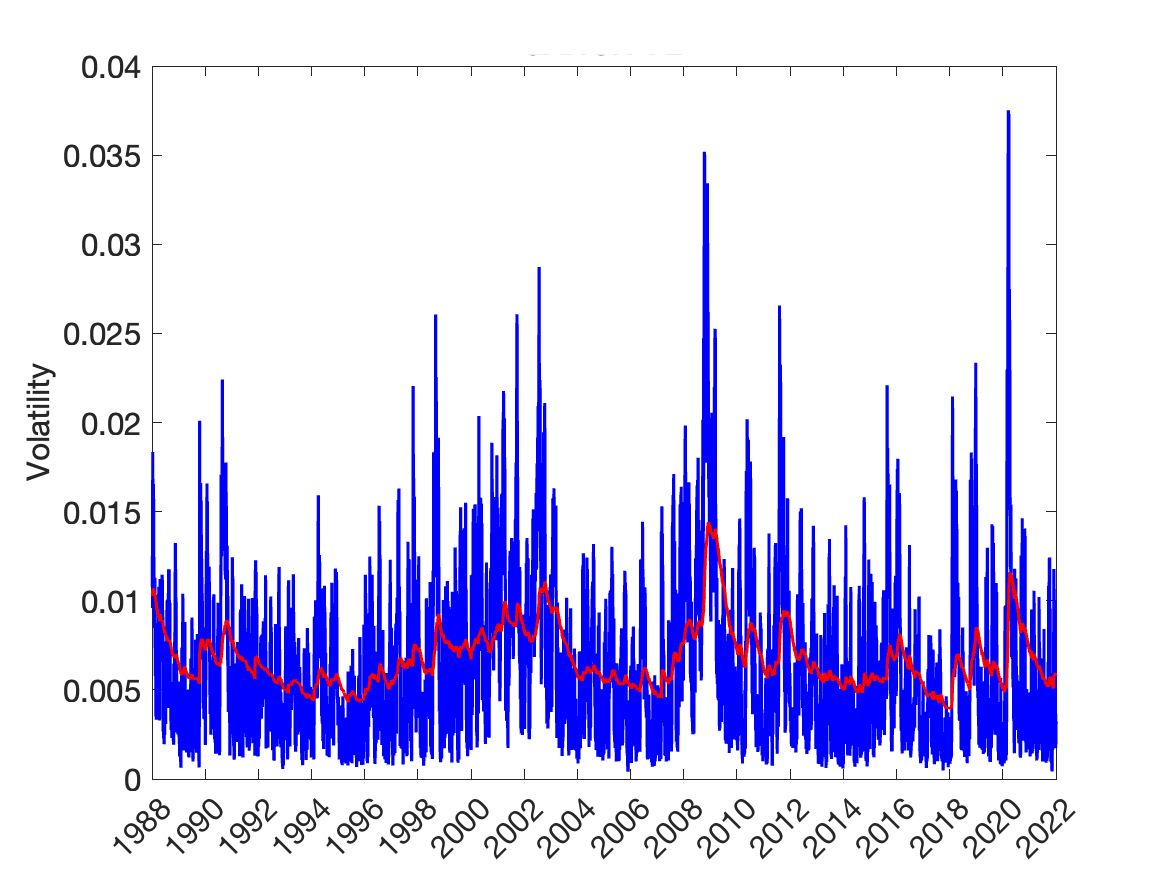}
\caption{Filtered conditional volatility for the $\textit{GARCH-HN}$ (left); $\textit{GARCH-CJOW}$ total volatility $\sqrt{q_t + s_t}$ (center blue) and long-term volatility component $\sqrt{q_t}$ (center red); $\textit{GARCH-2F}$ first volatility component $\sqrt{v_{1,t}}$ (right red) and second volatility component $\sqrt{v_{2,t}}$ (right blue).}
\label{fig:figure4}
\end{figure}

In Figure \ref{fig:figure4}, we plot the filtered conditional volatility of the $\textit{GARCH-HN}$ model and the two volatility components of the $\textit{GARCH-CJOW}$ model and $\textit{GARCH-2F}$ model. One of the volatility components of the $\textit{GARCH-2F}$ model mimics the dynamics of the one-factor $\textit{GARCH-HN}$, and the other volatility component is a slower moving process with much lower variation over the considered time span.

\begin{figure}[htbp]
\centering
\includegraphics[width=.5\textwidth]{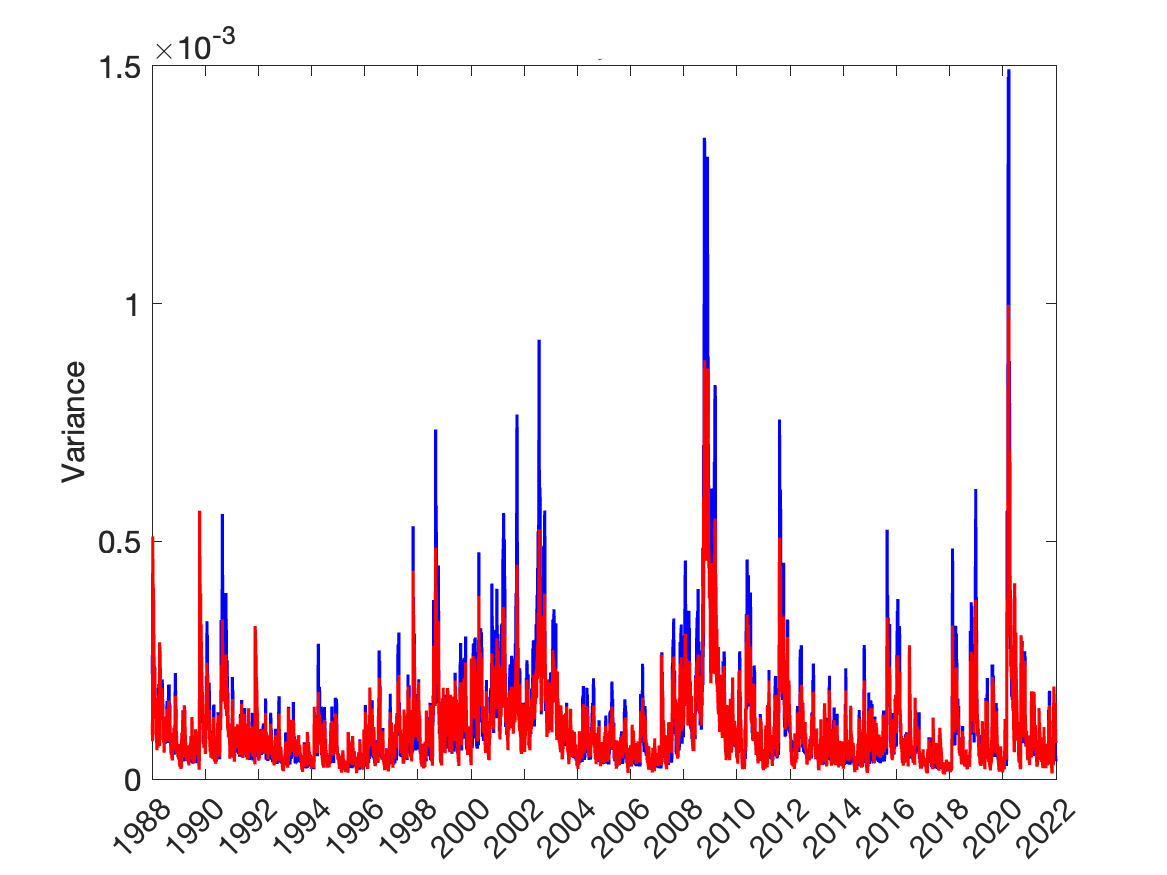}\hfill
\caption{$\textit{GARCH-HN}$ conditional variance (red) and $\textit{GARCH-2F}$ total conditional variance $v_{1,t} + v_{2,t}$ (blue).}
\label{fig:figure5}
\end{figure}

In Figure \ref{fig:figure5}, we compare the conditional variances of the one-factor $\textit{GARCH-HN}$ model and of the two-factor $\textit{GARCH-2F}$ model, that is $v_{1,t} + v_{2,t}$. The variance of both models displays a similar pattern, however during highly volatile periods, as in 2008 and 2020, the higher spikes of the $\textit{GARCH-2F}$  might yield a better descprition of the high volatility of such periods.

\subsection{Out-of-Sample Analysis}

To test the forecasting performances of the rival models, we forecast the Value-at-Risk ($\mathrm{VaR}$) at the 1\% and 5\% significance levels from one to five days ahead. We test the models during different periods of crisis by splitting the dataset into two subperiods (4534 observations each) and running separate out-of-sample analyses: from January 5, 1988 to December 21, 2005 (to include the September 11, 2001 crisis) and from December 22, 2005 to December 29, 2023 (to include the 2008 financial crisis and the COVID-19 period).
For each of the considered subsamples, we estimate the model parameters using a rolling window equal to 60\% of the days in the period. That is, we start by using the first 60\% of the daily S\&P500 total returns to calibrate the models, leaving the remaining data for the out-of-sample analysis. Then, for each day in the out-of-sample, we re-estimate the model parameters using the past 60\% daily observations and compute forecasts from one to five days ahead.

In particular, for all the competing models, at each forecasting time, we simulate $100,000$ sample paths of $R_{t+\ell}$, with $\ell = 1,2,3,4,5$, according to the return equation \eqref{eq:1}. This is done by randomly simulating the $Z_{1,t}$ and $Z_{2,t}$ standard normal random variables. Then, we estimate the $\mathrm{VaR}$ at the 1\% and 5\% significance levels by computing the empirical quantiles of the simulated sample paths of $R_{t+\ell}$.

We then measure the statistical accuracy of the VaR forecasts by performing the proportion of failures test of \cite{kupiec95} and the correct conditional coverage test of \cite{christoffersen98}. The results obtained for the proposed $\textit{GARCH-2F}$, $\textit{GARCH-2F}\alpha$ and  $\textit{GARCH-2F}\beta$, together with the competing $\textit{GARCH-2F}\alpha \beta$, $\textit{GARCH-HN}$ and $\textit{GARCH-CJOW}$ models, are reported in Tables \ref{tab:tab_outsample1} and \ref{tab:tab_outsample2}.

We note that, overall, the novel two-factor specifications result in a significantly lower number of failures compared to the one-factor models. For the period from January 1988 to December 2005, both the Kupiec and Christoffersen tests confirm the superior performance of the \textit{GARCH-2F} models, particularly the \textit{GARCH-2F}$\beta$, which consistently performs well at both the 95\% and 99\% confidence levels, outperforming the \textit{GARCH-2F}$\alpha \beta$ model with no spillovers between the volatility components. For the period from December 2005 to December 2023, the \textit{GARCH-2F} and \textit{GARCH-2F}$\beta$ models exhibit similar performances.

In addition, we further compare the accuracy of the VaR forecast using the Model Confidence Set (MCS) introduced by \cite{hansen11}. Following \cite{gonzales04}, we measure performance by means of the quantile loss function as in \cite{koenker78}, defined as

\begin{equation*}
    \mathrm{QL}_{t + \ell} (\vartheta)= (\vartheta - \mathds{1}_{ \{ R_{t + \ell}  < \mathrm{VaR}_{t+\ell}(\vartheta)\} }) (R_{t + \ell}  - \mathrm{VaR}_{t+\ell}(\vartheta)),
\end{equation*}

\noindent where $\mathrm{VaR}_{t+\ell}(\vartheta)$ denotes the $\mathrm{VaR}$ on day $t+\ell$ at the significance level $\vartheta$. Moreover, to check if the forecasts of the models $\textit{GARCH-CJOW}$, $\textit{GARCH-2F}$, $\textit{GARCH-2F}\alpha$, $\textit{GARCH-2F}\beta$ and $\textit{GARCH-2F}\alpha \beta$ significantly differ from the ones of the $\textit{GARCH-HN}$ benchmark we  performed the test of \cite{diebold02}. The quantile losses and the p-values are reported in Table \ref{tab:tab_outsample3}. 

Overall, the $\textit{GARCH-2F}$ model, its three nested specifications, and the $\textit{GARCH-CJOW}$ model return significantly different forecasts compared to the ones of the $\textit{GARCH-HN}$ as indicated by the p-values for both periods and significance levels. For the period from January 5, 1988 to December 21, 2005, all the four $\textit{GARCH-2F}$ models return lower quantile lossess, thus indicating a superior VaR forecasting accuracy and are included in the MCS, contrary to the $\textit{GARCH-CJOW}$ model. In the period from December 22, 2005 to December 29, 2023, only the $\textit{GARCH-2F}\beta$ is always included in the MCS and outperform the benchmark models by returning lower quantile losses.

\begin{table}[h!]
    \centering
\sisetup{table-number-alignment = center, % <-- added/changed
         table-space-text-pre ={(},
         table-space-text-post={\textsuperscript{***}},
         input-open-uncertainty={[},
         input-close-uncertainty={]},
         table-align-text-pre=false,
         table-align-text-post=false}
\begin{threeparttable}
    \caption{Likelihood ratio test statistic of \cite{kupiec95}, p-values in parentheses.}
    \label{tab:tab_outsample1}
\begin{tabular}{l 
                S[table-format=1.3] % <-- adopted to number of digits in numbers in cells
                S[table-format=1.3] % <-- adopted ...
                S[table-format=1.3] % <-- adopted ...
                S[table-format=1.3] % <-- adopted ..
                S[table-format=1.3] % <-- adopted ..
                 }
\toprule 
& \multicolumn{5}{c}{January 1988 to December 2005}            \\
\cmidrule(lr){2-6}
& \multicolumn{5}{c}{$\vartheta = 5\%$}            \\
\cmidrule(lr){2-6}
 & {$\ell = 1$} &  {$\ell = 2$} & {$\ell = 3$} & {$\ell = 4$} & {$\ell = 5$} \\
 \midrule
{\textit{HN}}   &  \fv{52.206}{0.000} & \fv{52.206}{0.000} & \fv{54.805}{0.000} & \fv{53.499}{0.000} & \fv{54.805}{0.000} \\ 
{\textit{CJOW} }   & \fv{36.641}{0.000}  & \fv{40.024}{0.000} & \fv{37.755}{0.000} & \fv{34.457}{0.000} & \fv{35.542}{0.000} \\
{\textit{2F}} & \fv{7.113}{0.007} & \fv{6.496}{0.011} & \fv{7.761}{0.005} & \fv{7.113}{0.007} & \fv{7.113}{0.007} \\
{\textit{2F$\beta$}} & \fv{2.275}{0.132} & \fv{1.644}{0.200} & \fv{1.644}{0.200} & \fv{1.644}{0.200} & \fv{1.946}{0.163} \\
{\textit{2F$\alpha$} } & \fv{7.415}{0.007} & \fv{7.965}{0.005} & \fv{9.120}{0.003} & \fv{9.724}{0.002} & \fv{13.008}{0.000} \\
{\textit{2F$\alpha \beta$} } & \fv{0.984}{0.321} & \fv{1.440}{0.230} & \fv{1.202}{0.273} & \fv{1.202}{0.273} & \fv{1.202}{0.273} \\
\midrule
& \multicolumn{5}{c}{December 2005 to December 2023}            \\
\cmidrule(lr){2-6}
& \multicolumn{5}{c}{$\vartheta =  5\% $}            \\
\cmidrule(lr){2-6}
 & {$\ell = 1$} &  {$\ell = 2$} & {$\ell = 3$} & {$\ell = 4$} & {$\ell = 5$} \\
\midrule
{\textit{HN}}   &  \fv{9.688}{0.001} & \fv{10.309}{0.001} & \fv{9.688}{0.001} & \fv{10.309}{0.001} & \fv{9.688}{0.001} \\ 
{\textit{CJOW} }   & \fv{2.264}{0.132} & \fv{2.923}{0.087} & \fv{2.923}{0.087} & \fv{2.264}{0.132} & \fv{2.583}{0.108}  \\
{\textit{2F}} & \fv{5.379}{0.020} & \fv{5.379}{0.020} & \fv{5.379}{0.020} & \fv{4.852}{0.021} & \fv{4.354}{0.037} \\
{\textit{2F$\beta$}} & \fv{1.961}{0.161} & \fv{2.648}{0.104} & \fv{2.648}{0.104} & \fv{2.291}{0.130} & \fv{2.648}{0.104} \\
{\textit{2F$\alpha$} } & \fv{2.648}{0.104} & \fv{2.648}{0.104} & \fv{2.648}{0.104} & \fv{1.961}{0.161} & \fv{1.961}{0.161}\\
{\textit{2F$\alpha \beta$} } & \fv{0.034}{0.854} & \fv{0.006}{0.940} & \fv{0.006}{0.940} & \fv{0.006}{0.940} & \fv{0.006}{0.940} \\
\midrule
& \multicolumn{5}{c}{January 1988 to December 2005}            \\
\cmidrule(lr){2-6}
& \multicolumn{5}{c}{$\vartheta = 1\%$}            \\
\cmidrule(lr){2-6}
 & {$\ell = 1$} &  {$\ell = 2$} & {$\ell = 3$} & {$\ell = 4$} & {$\ell = 5$} \\
 \midrule
{\textit{HN}}   &  \fv{81.442}{0.000} & \fv{73.481}{0.000} & \fv{76.104}{0.000} & \fv{73.481}{0.000} & \fv{70.891}{0.000}  \\ 
{\textit{CJOW}}   & \fv{115.931}{0.000} & \fv{115.931}{0.000} & \fv{98.179}{0.000} & \fv{95.317}{0.000} & \fv{89.679}{0.000} \\
{\textit{2F}} & \fv{0.781}{0.377} & \fv{0.437}{0.509} & \fv{0.001}{0.976} & \fv{0.001}{0.976} & \fv{0.042}{0.838} \\
{\textit{2F$\beta$}} & \fv{2.352}{0.125} & \fv{2.352}{0.125} & \fv{2.352}{0.125} & \fv{2.352}{0.125} & \fv{1.743}{0.187}  \\
{\textit{2F$\alpha$} } & \fv{0.001}{0.976} & \fv{0.073}{0.788} & \fv{0.263}{0.608} & \fv{0.073}{0.788} & \fv{0.073}{0.788}  \\
{\textit{2F$\alpha \beta$} } &  \fv{7.611}{0.006} & \fv{4.654}{0.031} & \fv{3.042}{0.081} & \fv{3.042}{0.081} & \fv{3.042}{0.081} \\
\midrule
& \multicolumn{5}{c}{December 2005 to December 2023}            \\
\cmidrule(lr){2-6}
& \multicolumn{5}{c}{$\vartheta = 1\%$}            \\
\cmidrule(lr){2-6}
 & {$\ell = 1$} &  {$\ell = 2$} & {$\ell = 3$} & {$\ell = 4$} & {$\ell = 5$} \\
 \midrule
{\textit{HN}}   &  \fv{49.055}{0.000} & \fv{40.323}{0.000} & \fv{38.239}{0.000} & \fv{38.239}{0.000} & \fv{38.239}{0.000} \\ 
{\textit{CJOW}}   & \fv{26.628}{0.000} & \fv{21.439}{0.000} & \fv{19.808}{0.000} & \fv{18.228}{0.000} & \fv{15.226}{0.000} \\
{\textit{2F}} & \fv{0.266}{0.606} & \fv{0.583}{0.445} & \fv{0.583}{0.445} & \fv{1.632}{0.201} & \fv{1.632}{0.201} \\
{\textit{2F$\beta$}} & \fv{5.558}{0.018} & \fv{3.801}{0.051} & \fv{3.033}{0.082} & \fv{3.033}{0.082} & \fv{2.344}{0.126} \\
{\textit{2F$\alpha$} } & \fv{15.226}{0.000} & \fv{12.445}{0.000} & \fv{11.141}{0.001} & \fv{9.897}{0.002} & \fv{8.715}{0.003}  \\
{\textit{2F$\alpha \beta$} } &  \fv{16.700}{0.000} & \fv{12.445}{0.000} & \fv{8.715}{0.003} & \fv{6.543}{0.011} & \fv{5.558}{0.018} \\
\bottomrule
\end{tabular}
\end{threeparttable}
\end{table}

\begin{table}[h!]
    \centering
\sisetup{table-number-alignment = center, % <-- added/changed
         table-space-text-pre ={(},
         table-space-text-post={\textsuperscript{***}},
         input-open-uncertainty={[},
         input-close-uncertainty={]},
         table-align-text-pre=false,
         table-align-text-post=false}
\begin{threeparttable}
   \caption{Likelihood ratio test statistic of \cite{christoffersen98}, p-values in parentheses.}
    \label{tab:tab_outsample2}
\begin{tabular}{l 
                S[table-format=1.3] % <-- adopted to number of digits in numbers in cells
                S[table-format=1.3] % <-- adopted ...
                S[table-format=1.3] % <-- adopted ...
                S[table-format=1.3] % <-- adopted ..
                S[table-format=1.3] % <-- adopted ..
                 }
\toprule 
& \multicolumn{5}{c}{January 1988 to December 2005}            \\
\cmidrule(lr){2-6}
& \multicolumn{5}{c}{$\vartheta = 5\%$}            \\
\cmidrule(lr){2-6}
 & {$\ell = 1$} &  {$\ell = 2$} & {$\ell = 3$} & {$\ell = 4$} & {$\ell = 5$} \\
 \midrule
{\textit{HN}}   &  \fv{53.399}{0.000} & \fv{53.399}{0.000} & \fv{55.774}{0.000} & \fv{54.577}{0.000} & \fv{56.362}{0.000} \\ 
{\textit{CJOW} }   & \fv{37.093}{0.000}  & \fv{41.201}{0.000} & \fv{39.923}{0.000} & \fv{ 37.138}{0.000} & \fv{40.089}{0.000} \\
{\textit{F2}} & \fv{9.294}{0.009} & \fv{8.515}{0.014} & \fv{10.111}{0.006} & \fv{9.294}{0.009} & \fv{13.271}{0.001} \\
{\textit{2F$\beta$}} & \fv{3.143}{0.208} & \fv{2.326}{0.313} & \fv{2.326}{0.313} & \fv{2.326}{0.313} & \fv{2.718}{0.257} \\
{\textit{2F$\alpha$} } & \fv{7.462}{0.024} & \fv{8.211}{0.017} & \fv{9.688}{0.008} & \fv{11.661}{0.003} & \fv{18.748}{0.000} \\
{\textit{2F$\alpha \beta$} } & \fv{1.030}{0.598} & \fv{2.341}{0.310} & \fv{3.154}{0.207} & \fv{3.154}{0.207} & \fv{3.154}{0.207} \\
\midrule
& \multicolumn{5}{c}{December 2005 to December 2023}            \\
\cmidrule(lr){2-6}
& \multicolumn{5}{c}{$\vartheta =  5\% $}            \\
\cmidrule(lr){2-6}
 & {$\ell = 1$} &  {$\ell = 2$} & {$\ell = 3$} & {$\ell = 4$} & {$\ell = 5$} \\
\midrule
{\textit{HN}}   &  \fv{11.631}{0.003} & \fv{12.103}{0.002} & \fv{11.631}{0.003} & \fv{12.103}{0.002} & \fv{11.631}{0.003} \\ 
{\textit{CJOW} }   & \fv{3.662}{0.160} & \fv{5.028}{0.081} & \fv{5.028}{0.081} & \fv{4.698}{0.096} & \fv{4.850}{0.089} \\
{\textit{2F}} & \fv{8.689}{0.013} & \fv{8.689}{0.013} & \fv{8.689}{0.013} & \fv{7.957}{0.019} & \fv{7.262}{0.026} \\
{\textit{2F$\beta$}} & \fv{3.839}{0.147} & \fv{4.841}{0.089} & \fv{4.841}{0.089} & \fv{5.878}{0.053} & \fv{6.456}{0.040} \\
{\textit{2F$\alpha$} } & \fv{8.416}{0.015} & \fv{8.416}{0.015} & \fv{8.416}{0.015} & \fv{7.176}{0.028} & \fv{7.176}{0.028} \\
{\textit{2F$\alpha \beta$} } & \fv{2.774}{0.250} & \fv{2.566}{0.277} & \fv{2.566}{0.277} & \fv{2.566}{0.277} & \fv{2.566}{0.277} \\
\midrule
& \multicolumn{5}{c}{January 1988 to December 2005}             \\
\cmidrule(lr){2-6}
& \multicolumn{5}{c}{$\vartheta = 1\%$}            \\
\cmidrule(lr){2-6}
 & {$\ell = 1$} &  {$\ell = 2$} & {$\ell = 3$} & {$\ell = 4$} & {$\ell = 5$} \\
 \midrule
{\textit{HN}}   &  \fv{82.208}{0.000} & \fv{74.554}{0.000} & \fv{77.067}{0.000} & \fv{74.554}{0.000} & \fv{72.078}{0.000} \\ 
{\textit{CJOW}}   & \fv{115.996}{0.000} & \fv{115.996}{0.000} & \fv{98.494}{0.000} & \fv{95.692}{0.000} & \fv{91.254}{0.000} \\
{\textit{2F}} & \fv{1.322}{0.516} & \fv{0.929}{0.628} & \fv{0.362}{0.834} & \fv{0.362}{0.834} & \fv{0.444}{0.801} \\
{\textit{2F$\beta$}} & \fv{3.206}{0.201} & \fv{3.206}{0.201} & \fv{3.051}{0.218} & \fv{3.051}{0.218} & \fv{2.387}{0.303}\\
{\textit{2F$\alpha$} } & \fv{0.362}{0.834} & \fv{0.395}{0.821} & \fv{0.548}{0.760} & \fv{0.395}{0.821} & \fv{0.395}{0.821}  \\
{\textit{2F$\alpha \beta$} } &  \fv{10.126}{0.006} & \fv{5.219}{0.074} & \fv{3.791}{0.150} & \fv{3.791}{0.150} & \fv{3.791}{0.150} \\
\midrule
& \multicolumn{5}{c}{December 2005 to December 2023}          \\
\cmidrule(lr){2-6}
& \multicolumn{5}{c}{$\vartheta = 1\%$}            \\
\cmidrule(lr){2-6}
 & {$\ell = 1$} &  {$\ell = 2$} & {$\ell = 3$} & {$\ell = 4$} & {$\ell = 5$} \\
 \midrule
{\textit{HN}}   &  \fv{49.385}{0.000} & \fv{40.478}{0.000} & \fv{38.360}{0.000} & \fv{38.360}{0.000} & \fv{38.360}{0.000}  \\ 
{\textit{CJOW}}   & \fv{26.633}{0.000} & \fv{21.445}{0.000} & \fv{19.824}{0.000} & \fv{18.258}{0.000} & \fv{16.768}{0.000} \\
{\textit{2F}} & \fv{2.558}{0.278} & \fv{3.111}{0.211} & \fv{3.111}{0.211} & \fv{1.820}{0.403} & \fv{1.820}{0.403} \\
{\textit{2F$\beta$}} & \fv{6.043}{0.049} & \fv{4.454}{0.108} & \fv{3.783}{0.151} & \fv{3.783}{0.151} & \fv{3.199}{0.202} \\
{\textit{2F$\alpha$} } & \fv{19.300}{0.000} & \fv{17.065}{0.000} & \fv{16.054}{0.000} & \fv{15.116}{0.001} & \fv{14.254}{0.001}  \\
{\textit{2F$\alpha \beta$} } &  \fv{16.750}{0.000} & \fv{12.586}{0.002} & \fv{8.999}{0.012} & \fv{6.954}{0.031} & \fv{6.043}{0.049} \\
\bottomrule
\end{tabular}
\end{threeparttable}
\end{table}

%qlvar

\begin{table}[h!]
%\footnotesize 
\resizebox{\textwidth}{!}{
    \centering
\sisetup{table-number-alignment = center, % <-- added/changed
         table-space-text-pre ={(},
         table-space-text-post={\textsuperscript{***}},
         input-open-uncertainty={[},
         input-close-uncertainty={]},
         table-align-text-pre=false,
         table-align-text-post=false}
\begin{threeparttable}
    \caption{ $\mathrm{QL}_{t + \ell} (\vartheta)$ ratios with respect to \textit{GARCH-HN} and p-values of the test of \cite{diebold02} are reported in parentheses (left column);  Models included in the MCS of \cite{hansen11} according to $\mathrm{QL}_{t+\ell}(\vartheta)$ function (right column).}
    \label{tab:tab_outsample3}
\begin{tabular}{l 
                S[table-format=-1.3] % <-- adopted to number of digits in numbers in cells
                S[table-format=-1] % <-- adopted to number of digits in numbers in cells
                S[table-format=-1.3] % <-- adopted to number of digits in numbers in cells
                S[table-format=-1] 
                S[table-format=-1.3] 
                S[table-format=-1] 
                S[table-format=-1.3] 
                S[table-format=-1] 
                S[table-format=-1.3] 
                S[table-format=-1] 
                 }
\toprule 
& \multicolumn{10}{c}{January 1988 to December 2005}            \\
\cmidrule(lr){2-11}
& \multicolumn{10}{c}{$\vartheta = 5\%$}            \\
\cmidrule(lr){2-11}
 & \multicolumn{2}{c}{$\ell = 1$} & \multicolumn{2}{c}{$\ell = 2$} & \multicolumn{2}{c}{$\ell = 3$} & \multicolumn{2}{c}{$\ell = 4$} & \multicolumn{2}{c}{$\ell = 5$} \\
\cmidrule(lr){2-3}   \cmidrule(lr){4-5} \cmidrule(lr){6-7} \cmidrule(lr){8-9} \cmidrule(lr){10-11}
 & {QL} & {MCS}   & {QL} & {MCS} & {QL} & {MCS} & {QL} & {MCS} & {QL} & {MCS}  \\
 \midrule
 {\textit{HN}} & 1 & {} & 1 & {} & 1 & {} & 1 & {} & 1 & {} \\
{\textit{CJOW}}   & \fv{1.021}{0.000}  & {} &\fv{1.023}{0.000} & {} & \fv{1.025}{0.000} & {} & \fv{1.025}{0.000} & {} & \fv{1.025}{0.000} & \\
{\textit{2F}} & \fv{1.073}{0.000} & {\checkmark} & \fv{1.074}{0.000} & {\checkmark} & \fv{1.075}{0.000} & {\checkmark} & \fv{1.077}{0.000} & {\checkmark} & \fv{1.078}{0.000} & {\checkmark}\\
{\textit{2F$\beta$}} & \fv{1.059}{0.000} & {\checkmark} & \fv{1.060}{0.000} & {\checkmark} & \fv{1.060}{0.000} & {\checkmark} & \fv{1.061}{0.000} & {\checkmark} & \fv{1.062}{0.000} & {\checkmark} \\
{\textit{2F$\alpha$} }& \fv{1.050}{0.000} & {\checkmark} & \fv{1.050}{0.000} & {\checkmark} & \fv{1.050}{0.000} & {\checkmark} & \fv{1.049}{0.000} & {\checkmark} & \fv{1.048}{0.000} & {\checkmark} \\
{\textit{2F$\alpha \beta$} } & \fv{1.056}{0.000} & {\checkmark} & \fv{1.056}{0.000} & {\checkmark} & \fv{1.057}{0.000} & {\checkmark} & \fv{1.057}{0.000} & {\checkmark} & \fv{1.058}{0.000} & {\checkmark} \\
\midrule
& \multicolumn{10}{c}{December 2005 to December 2023}            \\
\cmidrule(lr){2-11}
& \multicolumn{10}{c}{$\vartheta = 5\%$}            \\
\cmidrule(lr){2-11}
 & \multicolumn{2}{c}{$\ell = 1$} & \multicolumn{2}{c}{$\ell = 2$} & \multicolumn{2}{c}{$\ell = 3$} & \multicolumn{2}{c}{$\ell = 4$} & \multicolumn{2}{c}{$\ell = 5$} \\
\cmidrule(lr){2-3}   \cmidrule(lr){4-5} \cmidrule(lr){6-7} \cmidrule(lr){8-9} \cmidrule(lr){10-11}
 & {QL} & {MCS}   & {QL} & {MCS} & {QL} & {MCS} & {QL} & {MCS} & {QL} & {MCS}  \\
 \midrule
 {\textit{HN}} & 1 & {\checkmark} & 1 & {}  & 1 & {\checkmark} & 1 & {\checkmark} & 1 & {\checkmark} \\
{\textit{CJOW}} & \fv{1.015}{0.001} & {\checkmark} & \fv{1.017}{0.000} & {} & \fv{1.018}{0.000} & {\checkmark} & \fv{1.018}{0.000} & {\checkmark} & \fv{1.018}{0.000} & {\checkmark} \\
{\textit{2F}} & \fv{0.983}{0.000} & {} & \fv{0.986}{0.000} & {} & \fv{0.987}{0.000} & {} & \fv{0.988}{0.000} & {\checkmark} & \fv{0.988}{0.000} & {\checkmark} \\
{\textit{2F$\beta$}} & \fv{1.054}{0.000} & {\checkmark} & \fv{1.054}{0.000} & {\checkmark} & \fv{1.053}{0.000} & {\checkmark} & \fv{1.050}{0.000} & {\checkmark} & \fv{1.049}{0.000} & {\checkmark} \\
{\textit{2F$\alpha$} }& \fv{0.985}{0.000} & {} & \fv{0.985}{0.000} & {} & \fv{0.985}{0.000} & {} & \fv{0.985}{0.000} & {} & \fv{0.985}{0.000}  \\
{\textit{2F$\alpha \beta$} } &  \fv{1.027}{0.000} & {\checkmark} & \fv{1.028}{0.000} & {} & \fv{1.028}{0.000} & {\checkmark} & \fv{1.027}{0.000} & {\checkmark} & \fv{1.026}{0.000} & {\checkmark} \\
\midrule
& \multicolumn{10}{c}{January 1988 to December 2005}            \\
\cmidrule(lr){2-11}
& \multicolumn{10}{c}{$\vartheta = 1\%$}            \\
\cmidrule(lr){2-11}
 & \multicolumn{2}{c}{$\ell = 1$} & \multicolumn{2}{c}{$\ell = 2$} & \multicolumn{2}{c}{$\ell = 3$} & \multicolumn{2}{c}{$\ell = 4$} & \multicolumn{2}{c}{$\ell = 5$} \\
\cmidrule(lr){2-3}   \cmidrule(lr){4-5} \cmidrule(lr){6-7} \cmidrule(lr){8-9} \cmidrule(lr){10-11}
 & {QL} & {MCS}   & {QL} & {MCS} & {QL} & {MCS} & {QL} & {MCS} & {QL} & {MCS}  \\
 \midrule
 {\textit{HN}} & 1 &  & 1 &  & 1 &  & 1 &  & 1 & \\
 {\textit{CJOW}}  & \fv{0.919}{0.000} & {} & \fv{0.927}{0.000} & {} & \fv{0.932}{0.000} & {} & \fv{0.934}{0.000} & {} & \fv{0.936}{0.000} & {} \\
{\textit{2F}} & \fv{1.257}{0.000} & {\checkmark} & \fv{1.236}{0.000} & {\checkmark} & \fv{1.231}{0.000} & {\checkmark} & \fv{1.221}{0.000} & {\checkmark} & \fv{1.224}{0.000} & {\checkmark} \\
{\textit{2F$\beta$}} & \fv{1.244}{0.000} & {\checkmark} & \fv{1.234}{0.000} & {\checkmark} & \fv{1.233}{0.000} & {\checkmark} & \fv{1.223}{0.000} & {\checkmark} & \fv{1.221}{0.000} & {\checkmark} \\
{\textit{2F$\alpha$} } & \fv{1.240}{0.000} & {\checkmark} & \fv{1.229}{0.000} & {\checkmark} & \fv{1.221}{0.000} & {\checkmark} & \fv{1.210}{0.000} & {\checkmark} & \fv{1.201}{0.000} & {\checkmark} \\
{\textit{2F$\alpha \beta$} } & \fv{1.237}{0.000} & {\checkmark} & \fv{1.234}{0.000} & {\checkmark} & \fv{1.234}{0.000} & {\checkmark} & \fv{1.230}{0.000} & {\checkmark} & \fv{1.226}{0.000} & {\checkmark} \\
\midrule
& \multicolumn{10}{c}{December 2005 to December 2023}            \\
\cmidrule(lr){2-11}
& \multicolumn{10}{c}{$\vartheta = 1\%$}            \\
\cmidrule(lr){2-11}
 & \multicolumn{2}{c}{$\ell = 1$} & \multicolumn{2}{c}{$\ell = 2$} & \multicolumn{2}{c}{$\ell = 3$} & \multicolumn{2}{c}{$\ell = 4$} & \multicolumn{2}{c}{$\ell = 5$} \\
\cmidrule(lr){2-3}   \cmidrule(lr){4-5} \cmidrule(lr){6-7} \cmidrule(lr){8-9} \cmidrule(lr){10-11}
 & {QL} & {MCS}   & {QL} & {MCS} & {QL} & {MCS} & {QL} & {MCS} & {QL} & {MCS}  \\
 \midrule
 {\textit{HN}} & 1 &  & 1 &  & 1 &  & 1 &  & 1 & \\
{\textit{CJOW}}   & \fv{1.097}{0.000} & {\checkmark} & \fv{1.104}{0.000} & {\checkmark} & \fv{1.111}{0.000} & {\checkmark} & \fv{1.114}{0.000} & {\checkmark} & \fv{1.119}{0.000} & {\checkmark} \\
{\textit{2F} } & \fv{1.138}{0.001} & {\checkmark} & \fv{1.134}{0.001} & {\checkmark} & \fv{1.129}{0.001} & {\checkmark} & \fv{1.121}{0.001} & {\checkmark} & \fv{1.120}{0.001} & {\checkmark} \\
{\textit{2F$\beta$} } & \fv{1.127}{0.000} & {\checkmark} & \fv{1.131}{0.000} & {\checkmark} & \fv{1.129}{0.000} & {\checkmark} & \fv{1.132}{0.000} & {\checkmark} & \fv{1.134}{0.000} & {\checkmark} \\
{\textit{2F$\alpha$} } & \fv{1.101}{0.002} & {\checkmark} & \fv{1.099}{0.001} & {\checkmark} & \fv{1.096}{0.001} & {\checkmark} & \fv{1.095}{0.001} & {\checkmark} & \fv{1.094}{0.002} & {\checkmark} \\
{\textit{2F$\alpha \beta$} } & \fv{1.144}{0.007} & {\checkmark} & \fv{1.151}{0.003} & {\checkmark} & \fv{1.151}{0.002} & {\checkmark} & \fv{1.154}{0.001} & {\checkmark} & \fv{1.157}{0.001} & {\checkmark} \\
\bottomrule
\end{tabular}
\smallskip
\footnotesize
Values greater than one indicate better out-sample performance.
\end{threeparttable}
}
\end{table}

\section{Option Pricing} \label{sec_options}

In this section, we derive results that allow us to value option contracts using the four proposed two-factor models. First, we perform the risk-neutralization of these models. Second, we estimate the \textit{GARCH-2F} model (and its nested models) by jointly using returns and options data.

\subsection{ The \textit{GARCH-2F} Risk Neutral Dynamics}

Equations \eqref{eq:1}-\eqref{eq:2} define the return and the volatility dynamics under the physical measure. In this section, we risk-neutralize the model, so that we can compute option prices. Following \cite{christoffersen12}, we consider a pricing kernel with affine dynamic

\begin{equation} \label{eq_pricingk}
    M_{t} = \frac{e^{- r - \xi (\sqrt{{v}_{1,t}} Z_{1,t} + \sqrt{{v}_{2,t}} Z_{2,t} ) }}{\mathbb{E}[ e^{ - \xi (\sqrt{v_{1,t}} Z_{1,t} + \sqrt{v_{2,t}} Z_{2,t} ) } \mid \mathcal{F}_{t-1}]},
\end{equation}

\noindent where $\mathbb{E}$ denotes the expectation operator under the physical measure. Accordingly, we guess the conditional Radon–Nikodym derivative in the following form

\begin{equation} \label{eq_rn_derivative}
     \frac{dQ_{t}}{dP_{t}} = \frac{e^{  \Lambda ( \sqrt{{v}_{1,t}} Z_{1,t} + \sqrt{{v}_{2,t}} Z_{2,t})  } } {\mathbb{E} \left[e^{   \Lambda ( \sqrt{v_{1,t}} Z_{1,t} + \sqrt{v_{2,t}} Z_{2,t})  } \middle| \mathcal{F}_{t-1} \right]},
\end{equation}

\noindent and we impose

\begin{equation} \label{eq_rn_and_pricingk}
     \frac{dQ_{t}}{dP_{t}} = M_{t} e^r,
\end{equation}

\noindent so that using equations \eqref{eq_pricingk}-\eqref{eq_rn_and_pricingk} we obtain $\Lambda = - \xi$. Therefore, we have the following result.

\begin{prop} \label{prop_option1}
    Let us consider the return process specified by equation \eqref{eq:1}, then the equivalent martingale measure $\mathbb{Q}$ associated to the Radon–Nikodym derivative in \eqref{eq_rn_derivative} exists if and only if the coefficient $\Lambda$  satisfy 
    \begin{align} \label{eq_Lambda}
        \Lambda &= - \lambda - \frac{1}{2}.    
    \end{align}
\end{prop}

Finally, the following proposition yields the dynamic of the risk-neutralized processes $ \{R_t\}$,  $ \{v_{1,t}\}$ and $ \{v_{2,t}\}$.

\begin{prop} \label{prop_option2}
    Under the risk-neutral measure $\mathbb{Q}$, the dynamics of the return and of the two variance components are as follows:
    \begin{align}
    R_t &= r - \frac{1}{2} \left(v_{1,t} +  v_{2,t} \right) + \sqrt{v_{1,t}}Z^*_{1,t} + \sqrt{v_{2,t}}Z^*_{2,t}, \\
    \begin{bmatrix}
    v_{1,t+1} \\
    v_{2,t+1} 
    \end{bmatrix} &=
    \begin{bmatrix}
        \omega_1 \\
        \omega_2
    \end{bmatrix} + 
    \begin{bmatrix}
        \beta_{11} & \beta_{12} \\
        \beta_{21} & \beta_{22} 
    \end{bmatrix} 
    \begin{bmatrix}
    v_{1,t} \\
    v_{2,t} \\
    \end{bmatrix} + 
    \begin{bmatrix}
    \alpha_{11} & 
    \alpha_{12} \\
    \alpha_{21} & \alpha_{22} \\
    \end{bmatrix}
    \begin{bmatrix}
       \left(Z^*_{1,t}-\gamma^*_1 \sqrt{v_{1,t}} \right)^2 \\
       \left(Z^*_{2,t}-\gamma^*_2 \sqrt{v_{2,t}} \right)^2 
    \end{bmatrix},
    \end{align}
    where $Z^*_{1,t} \sim N(0,1) $ and $Z^*_{2,t} \sim N(0,1) $ denote the risk-neutral shocks and $\gamma^*_1 = \gamma_1 + \lambda + \frac{1}{2}$ and $\gamma^*_2 = \gamma_2 + \lambda + \frac{1}{2}$ are risk-neutralized parameters.
\end{prop}

\subsection{Moment Generating Function} \label{sec_mgf}

Equations \eqref{eq:1}-\eqref{eq:2} allow us to derive the moment generating function using a standard procedure as in \cite{heston00}. 

%Analytical formula for option prices allow for fast pricing but a shortcoming of the HNGARCH model is the assumption of normality which is at odds with observed empirical regularities for asset returns which often require skewed and fat-tailed conditional distributions. 

\begin{prop} \label{prop_option3}
     The conditional moment generating function for the logarithm of the terminal stock price $\ln{S_T}$, that is
    \begin{equation} \label{eq_momgenfun}
        f_t(\phi) = \mathbb{E} \left[ e^{\phi \ln{S_T}} \middle| \mathcal{F}_t \right], \quad t < T, 
    \end{equation}  
    
 \noindent can be computed as 
    \begin{equation}\label{eq_guess}
        f_t(\phi) = e^{\phi \ln S_t +  A_t(\phi) + B_{1,t}(\phi) {v}_{1, t+1}
        + B_{2,t}(\phi) {v}_{2, t+1}  },
    \end{equation}

\noindent where the expressions for $A_t(\phi)$, $B_{1,t}(\phi)$ and $B_{2,t}(\phi)$ can be computed using the recursion equations \eqref{eq_coeff_rico} in the Appendix.  

\end{prop}

\subsection{Valuation of Options}

Let us consider a European Call option on the underlying asset $S_t$ with strike price $K$ and maturity $T$. Using the inversion theorem in \cite{gilpelaez51}, the option price at a generic time $t < T$, which we denote by $C_t$, is computed as follows:

\begin{align} \label{eq_prezzo_opzione}
    C_t = \frac{1}{2} S_t &+ \frac{e^{-r(T-t)}}{\pi} \int_{0}^{\infty} \text{Re} \left[ \frac{ K^{-i \phi} f^*_t(i \phi + 1) }{ i \phi} \right] d\phi \\ \notag
    &- K e^{-r(T-t)} \left(\frac{1}{2} + \frac{1}{\pi}
    \int_{0}^{\infty} \text{Re} \left[ \frac{ K^{-i \phi} f^*_t(i \phi ) }{ i \phi } \right] d\phi  \right),
\end{align}

\noindent where, as in \cite{heston00},  $f^*_t$ denotes the (conditional) moment generating function under the risk-neutral measure, that is function \eqref{eq_guess} evaluated using risk-neutral parameters. Put option values can be calculated using the well-known put-call parity as in \cite{heston00}.

\section{Option Valuation Empirics} \label{sec_option_empirics}

We consider European options, both Put and Call, written on the S\&P500 index, with data retrieved
from Thomson Reuters Eikon Datastream. Specifically, we consider options with maturities ranging from 2021 to 2023 and the time series of their daily prices from February 10, 2021 to December 29, 2023.

As a common practice, see for example \cite{christoffersen12}, \cite{enzo23} and \cite{enzo24}, we apply several exclusion filters to obtain the final panel of option contracts. We keep only the options with time-to-maturity between 14 and 365 days and we select only out-of-the-money Put and Call options (we compute the moneyness as $K/S_t$, where $K$ is the strike price and $S_t$ is the underlying index level), and we filter out illiquid quotes by selecting only the six most liquid strikes at each maturity, and we consider option quotes only on Wednesday. Finally, we remove price quotes lower than $3.8\$$. 

In Table \ref{tab:optiondata}, we report the resulting number of option prices on the S\&P500 index, sorted by moneyness and days to maturity, for a total of $N=9372$ options prices.

\begin{table}[htbp]
\centering
\sisetup{table-number-alignment = center, % <-- added/changed
         table-space-text-pre ={(},
         table-space-text-post={\textsuperscript{***}},
         input-open-uncertainty={[},
         input-close-uncertainty={]},
         table-align-text-pre=false,
         table-align-text-post=false}
\begin{threeparttable}
    \caption{Number of S\&P500 option prices considered, sorted by moneyness and day-to-maturity (\textit{DTM}).}
    \label{tab:optiondata}
\begin{tabular}{l 
                S[table-format=-1.3] % <-- adopted to number of digits in numbers in cells
                S[table-format=-1] % <-- adopted to number of digits in numbers in cells
                S[table-format=-1.3] % <-- adopted to number of digits in numbers in cells
                S[table-format=-1] 
                 }
\toprule 
& \multicolumn{4}{c}{Number of option prices}            \\
\cmidrule(lr){2-5}
& \multicolumn{4}{c}{Maturity}            \\
\cmidrule(lr){2-5}
 {Moneyness} & {$14 \leq DTM \leq 50$} & {$50 < DTM \leq 150$} & {$150 < DTM \leq 365$} & {All}\\
 \midrule
{$0.8 \leq K/S_t \leq 0.9$} & 156 & 573 & 1133 & 1862\\
{$0.9 < K/S_t \leq 1.02$} & 817 & 1997 & 2527 & 5341\\
{$1.02 < K/S_t \leq 1.2$} & 245 & 807 & 1117 & 2169\\
{All} & 1218 & 3377 & 4777 & 9372\\
\bottomrule
\end{tabular}
\end{threeparttable}
\end{table}

\subsection{Parameter Estimation using both Returns and Options}

A common practice for estimating parameters using daily time series of log-returns and a panel of option contracts is using a suitably weighted log-likelihood function, see \cite{christoffersen12} and \cite{enzo23}. The log-likelihood for the returns series, $\ell_{\text{returns}}(\boldsymbol{\theta})$, is already available in equation \eqref{eq_likelihoodreturns}, whereas, for the $N$ option prices, we proceed as follows. First, we define the error on the volatility implied by the $i$-th option price as

\begin{equation}
    e_i = \frac{\textit{IV}^{MKT}_i-\textit{IV}^{MOD}_i}{\textit{IV}^{MKT}_i}, 
\end{equation}

\noindent where  $\textit{IV}^{MKT}_i$ and $\textit{IV}^{MOD}_i$ denote the market and the model volatilities implied by the $i$-th option price, which we compute via the Black-Scholes model. Then, by assuming a Gaussian distribution for $e_i$, we can derive the log-likelihood associated to the option pricing errors as

\begin{equation}
    \ell_{\text{options}}(\boldsymbol{\theta}) = -\frac{N}{2} \log{(2 \pi \sigma^2_e)} - \frac{1}{2} \sum_{i=1}^N \frac{e^2_i}{\sigma^2_e},
\end{equation}

\noindent where $\sigma^2_e$ represents the variance of the option pricing errors.

The number $N$ of data points available from the option panel could be significantly higher than the number $T$ of daily returns, see Section 4. Therefore, following  \cite{christoffersen12}, we assign an equal weight to returns and option prices by considering the following weighted joint (total) log-likelihood:

\begin{equation}
    \ell_{\text{joint}}(\boldsymbol{\theta}) = \frac{N+T}{2} \frac{ \ell_{\text{options}}(\boldsymbol{\theta})}{N} + \frac{N+T}{2} \frac{\ell_{\text{returns}}(\boldsymbol{\theta})}{T}.
\end{equation}

The joint estimation results are presented in Table \ref{tab:optionreturns}. 
We calculated standard information metrics AIC and BIC, and we also report the persistence values, as we did in Table \ref{tab_insample}. We observe different persistence values around 0.98 and 0.97 for the two volatility components, compared to the returns data-only estimation, which were 0.99 and 0.92. Additionally, we report the variance of the option pricing errors. The proposed \textit{GARCH-2F$\beta$} model return the highest joint likelihood values and lower AIC and BIC compared to the benchmark models, indicating improved performance in fitting both option and returns data. The \textit{GARCH-2F} models allowing for spillovers have highly significant parameters and provide a superior fit compared to the \textit{GARCH-2F$\alpha \beta$} model.

\begin{table}[htbp]
\centering
\begin{threeparttable}
    \caption{Joint maximum likelihood estimation results.}
    \label{tab:optionreturns}
\begin{tabular}{r 
                S[table-format=-3.3]
                S[table-format=-3.3]
                S[table-format=-3.3]
                S[table-format=-3.3]
                S[table-format=-3.3]
                S[table-format=-3.3]
                 }
\toprule
                    &   {\textit{GARCH-HN}}           &   {\textit{GARCH-CJOW} }      & {\textit{GARCH-2F} }     &  {\textit{GARCH-2F$\beta$} }             &  {\textit{GARCH-2F$\alpha$} } &  {\textit{GARCH-2F$\alpha \beta$} } \\
\midrule
$\omega_1$         &  {6.529e-07}\tnote{***} & {8.109e-06}\tnote{***} & {1.207e-12} & {1.207e-12} & {1.207e-12} & {1.207e-12} \\
                  & {(2.828e-11)} &  {(3.053e-08)} &  {(3.195e-12)} &  {(1.624e-12)} &  {(1.804e-11)} &  {(1.898e-12)}  \\
\addlinespace
$\omega_2$          &        &  & {1.207e-12} & {1.207e-12} & {4.928e-08}\tnote{***} & {4.928e-08}\tnote{***} \\
                  &  &   & {(2.887e-12)} &  {(4.240e-12)} &  {(4.081e-11)} &  {(6.855e-12)}   \\
\addlinespace
$\alpha_{11}$       &   {1.738e-06}\tnote{***} & {4.972e-06}\tnote{***} & {3.545e-11} & {3.545e-11}\tnote{***} & {3.545e-11} & {3.545e-11} \\
                  & {(2.279e-10)} &  {(7.820e-09)} & {(3.411e-12)}  &  {(3.532e-12)} &   {(5.671e-12)} &  {(2.660e-12)}   \\
\addlinespace
$\gamma_{1}$        &   {335.931}\tnote{***} & {291.608}\tnote{***} & {132.891}\tnote{***} & {132.892}\tnote{***} & {132.891}\tnote{***} & {132.891}\tnote{***} \\
                  & {(0.016)} &  {(0.443)} &  {(0.043)} &   {(0.091)} &  {(0.028)} &  {(0.014)}   \\
\addlinespace
$\beta_{11}$        &   {0.772}\tnote{***} & {0.907}\tnote{***} & {0.985}\tnote{***} & {0.985}\tnote{***} & {0.984}\tnote{***} & {0.985}\tnote{***} \\
                  & {(1.337e-04)} &  {(0.002)} &  {(1.107e-04)} &  {(5.428e-04)} &  {(1.617e-04)} &  {(2.719e-04)}   \\
\addlinespace
$\alpha_{22}$       &                         & {2.525e-06}\tnote{***} & {2.175e-06}\tnote{***} & {2.175e-06}\tnote{***} & {2.175e-06}\tnote{***} & {2.168e-06}\tnote{***} \\
                  &  &  {(4.856e-07)} & {(2.618e-10)} &  {(4.533e-10)} &  {(1.704e-09)} &  {(6.660e-10)} \\
\addlinespace
$\gamma_{2}$        &                     & {149.761}\tnote{***} & {274.319}\tnote{***} & {274.323}\tnote{***} & {274.319}\tnote{***} & {274.318}\tnote{***} \\
                  &  &  {(0.017)} &  {(0.087)} &  {(0.434)} &  {(0.064)} &  {(0.072)}   \\
\addlinespace
$\beta_{22}$        &                          & {0.741}\tnote{***} & {0.803}\tnote{***} & {0.802}\tnote{***} & {0.803}\tnote{***} & {0.804}\tnote{***} \\
                  &  & {(5.498e-04)} & {(1.007e-04)} &  {(3.695e-04)} &  {(8.350e-05)} &  {(2.134e-04)}  \\
\addlinespace
$\alpha_{12}$       &                         &  & {1.135e-11}\tnote{***} & {1.135e-11}\tnote{***} &  &  \\
                  &  &   & {(1.475e-12)} &  {(1.255e-12)} &   &   \\
\addlinespace
$\alpha_{21}$      &                         &  & {1.230e-13} & {1.231e-13} &  &  \\
                  &  &   & {(1.965e-12)} &  {(3.158e-12)} &   &    \\
\addlinespace
$\beta_{12}$        &                         &  & {0.000} &  & {1.622e-05}\tnote{***} &  \\
                  &  &   & {(4.466e-08)} &   &  {(1.350e-08)} &    \\
\addlinespace
$\beta_{21}$        &                         &  & {3.220e-05}\tnote{***} &  & {1.622e-05}\tnote{***} &  \\
                  &  &   & {(4.902e-09)} &   &  {(1.414e-08)} &    \\
\addlinespace
$\lambda$       &  {0.158}\tnote{***}      & {-6.816}\tnote{***} & {3.179}\tnote{***} & {3.177}\tnote{***} & {3.179}\tnote{***} & {3.178}\tnote{***} \\
                  & {(1.583e-05)} &  {(0.007)} &   {(5.890e-04)} &  {(0.002)} &  {(0.004)} &  {(3.443e-04)}   \\
\midrule
Log-likelihood      & {28,039}      &  {27,251}  &      {28,135}         & {28,135}    &    {28,126} & {28,118}     \\
AIC & {-56,068} & {-54,486} & {-56,244} & {-56,248} & {-56,230} & {-56,218} \\
BIC & {-56,029} & {-54,443} &  {-56,142} & {-56,162} & {-56,144} & {-56,148} \\ 
$\sigma^2_e$      & 0.084 & 0.076 & 0.085 & 0.083 & 0.085 & 0.086 \\          
Persistence   & 0.97   &     {0.96 $\vert$ 0.81}  & {0.98 $|$ 0.97}  &  {0.98 $|$ 0.97}   &  {0.98 $|$ 0.97}  & {0.98 $|$ 0.97}  \\
\bottomrule
\end{tabular} 
    \begin{tablenotes}
        \item The estimation period spans from January 5, 1988 to December 29, 2023 (9069 observations) for the returns, and spans from February 10, 2021 to November 29, 2023 (9372 observations) for the option prices. The standard errors, reported in parenthesis, are computed by inverting the negative Hessian matrix evaluated at the optimum parameter values. The variance $\sigma^2_e$ is calculated as the empirical variance estimator of the $N$ option implied volatility errors. [*],[**],[***] denote statistical significance at the 0.1, 0.05, 0.01 levels, respectively.
    \end{tablenotes}
\end{threeparttable}
\end{table}

\newpage

\subsection{Empirical Results for Options}

To assess the performance of the six competing models (including the models nested by the \textit{GARCH-2F}) in option pricing, we follow \cite{christoffersen12} and employ the (percentage) implied volatility root mean square error:

\begin{equation}
    \textit{IVRMSE(\%)} = 100 \times \sqrt{ \frac{1}{N} \sum_{i=1}^N \left( \textit{IV}^{MKT}_i - \textit{IV}^{MOD}_i \right)^2}.
\end{equation}

The results obtained, shown in Table \ref{tab:impliedvol}, reveal that the novel two-factor specifications yield better performance compared to the benchmark approaches when pricing options. In particular, the \textit{GARCH-2F} model and its nested models return lower (from 2\% to 4\%) overall implied volatility pricing errors compared to the one factor \textit{GARCH-HN} benchmark model. By sorting the options for moneyness and days to maturity, the \textit{GARCH-2F} and its nested models provide consistent results, that is lower \textit{IVRMSE} than the one-factor benchmark models, and reveal more robust performance in pricing out-of-the money options ($0.8 \leq K/S_t \leq 0.9$). Notably, the parsimonious \textit{GARCH-2F$\alpha\beta$} model demonstrates slightly better performance across both moneyness and days to maturity.

\begin{table}[htbp]
\centering
\sisetup{table-number-alignment = center, % <-- added/changed
         table-space-text-pre ={(},
         table-space-text-post={\textsuperscript{***}},
         input-open-uncertainty={[},
         input-close-uncertainty={]},
         table-align-text-pre=false,
         table-align-text-post=false}
\begin{threeparttable}
\caption{Implied volatility root mean square error, in percentage points.}
\label{tab:impliedvol}
\begin{tabular}{l
                S[table-format=-1.3] % <-- adopted to number of digits in numbers in cells
                S[table-format=-1.3] % <-- adopted to number of digits in numbers in cells
                S[table-format=-1.3] % <-- adopted to number of digits in numbers in cells
                S[table-format=-1.3] 
                S[table-format=-1.3] 
                S[table-format=-1.3] 
                 }
\toprule 
Panel A: \textit{IVRMSE(\%)} overall \\
\midrule
& {$\textit{HN}$} & {$\textit{CJOW}$} & {$\textit{F2}$} & {$\textit{F2}\beta$} & {$\textit{F2}\alpha$} & {$\textit{F2}\alpha\beta$}  \\
\midrule
& 5.383 & 5.573 & 5.273 & 5.273 & 5.200 & 5.161  \\
{Normalized} & 1.000 & 1.035 & 0.980 & 0.980 & 0.966 & 0.959 \\
\midrule
Panel B: \textit{IVRMSE(\%)} by moneyness \\
\midrule
{Moneyness} & {$\textit{HN}$} & {$\textit{CJOW}$} & {$\textit{F2}$} & {$\textit{F2}\beta$} & {$\textit{F2}\alpha$} & {$\textit{F2}\alpha\beta$}  \\
\midrule
{$0.8 \leq K/S_t \leq 0.9$} & 7.212 & 7.418 & 6.936 & 6.935 & 6.831 & 6.765 \\
{Normalized} & 1.000 & 1.029 & 0.962 & 0.962 & 0.947 & 0.938 \\
{$0.9 < K/S_t \leq 1.02$} &  4.918 & 5.091 & 4.835 & 4.834 & 4.769 & 4.735 \\
{Normalized} & 1.000 & 1.035 & 0.983 & 0.983 & 0.970 & 0.963 \\
{$1.02 < K/S_t \leq 1.2$} & 4.874 & 5.103 & 4.873 & 4.873 & 4.810 & 4.784 \\
{Normalized} & 1.000 &  1.047 & 1.000 & 1.000 & 0.987 & 0.982  \\
\midrule
Panel C: \textit{IVRMSE(\%)} by day-to-maturity (DTM) \\
\midrule
{Maturity} & {$\textit{HN}$} & {$\textit{CJOW}$} & {$\textit{F2}$} & {$\textit{F2}\beta$} & {$\textit{F2}\alpha$} & {$\textit{F2}\alpha\beta$}  \\
\midrule
{$14 \leq DTM \leq 50$} & 5.902 & 5.658 & 5.819 & 5.818 & 5.746 & 5.712 \\
{Normalized} & 1.000 &  0.959 & 0.986 & 0.986 & 0.974 & 0.968 \\
{$50 < DTM \leq 150$} & 5.050 & 5.200 & 4.949 & 4.948 & 4.872 & 4.832 \\
{Normalized} & 1.000 &  1.030 & 0.980 & 0.980 & 0.965 & 0.957 \\
{$150 < DTM \leq 365$} & 5.478 & 5.809 & 5.355 & 5.355 & 5.284 & 5.243 \\
{Normalized} & 1.000 & 1.061 & 0.978 & 0.978 & 0.965 & 0.957 \\
\bottomrule
\end{tabular}
\end{threeparttable} 
\end{table}

\section{Conclusions}\label{sec_conclusion}

Accurately capturing the the complex and multi-faceted nature of volatility is crucial for effective financial analysis and risk management. In this paper, we introduce a novel GARCH specification with two volatility components each driven by an independent stochastic factor, an approach commonly used in the continuous time (stochastic) volatility models, see, e.g.,  \cite{christoffersen08} and \cite{fouque11}, yet still overlooked in discrete-time GARCH models. 

%The proposed model, named $\textit{GARCH-2F}$, extends the work of \cite{heston00} by introducing two volatility components for the returns, each driven by its own source of randomness. Then, we contribute to the component GARCH literature by providing an extension to two stochastic innovations to better capture the multi-faceted nature of volatility.

We conducted a theoretical investigation that establishes, for the first time, sufficient conditions for strict stationarity and geometric ergodicity in a two-component GARCH model driven by independent stochastic factors. Additionally, we derived a continuous-time limit for the proposed \textit{GARCH-2F} model, which recovers a well-known bivariate square-root stochastic volatility specification. Finally, we considered three more parsimonious nested models: the \textit{GARCH-2F}$\alpha$, \textit{GARCH-2F}$\beta$, and \textit{GARCH-2F}$\alpha \beta$, the latter introduced by \cite{ghabani24}.

We empirically tested the two-factor models on S\&P500 total log-returns data and compare their in-sample and out-sample performance against one-factor affine benchmark models. Overall, the two-factor models demonstrate superior in-sample and out-sample $\mathrm{VaR}$ prediction accuracy. Then, we also considered the pricing of options on the S\&P500 index. The two-factors model provide lower implied volatility pricing errors, particularly for out-of-the money options, when estimated jointly using returns and options data.

The approach proposed in this article can be extended in many directions. Potential extensions include include incorporating fat-tailed distributions for the innovations while preserving model affinity, exploring different specifications for the drift of the return process (e.g., an autoregressive model), and considering additional lags.

\section*{Appendix}

\textit{Proof of Lemma \ref{lemma1}.} We need to prove that for each $k \in \mathbb{N}$, the $k$-step transition probability density of $\{ \boldsymbol{v}_t \}$ is strictly positive over the set of states $\Omega$ defined below. This ensures that if the set $\Omega$ has positive Lebesgue measure, then $\{ \boldsymbol{v}_t \}$ is irreducible w.r.t. the Lebesgue measure restricted to $\Omega$, that is the measure $\psi$, see \cite{tweedie93}.

We begin by studying the set of states that are reachable in one step from an initial state $\boldsymbol{x}=(x_1, x_2)^T \in \mathcal{D}$, which we denote by $G_{x,1}$. 

Given an initial value $\boldsymbol{x}$ and using equations \eqref{eq:2} and Assumption 1, we derive the set of points $\boldsymbol{y}=(y_1,y_2)^T$ that can be reached by the process $\{\boldsymbol{v}_t\}$ in one step starting from $\boldsymbol{x}$ as

\begin{equation*}
    G_{x,1} = \left\{ \boldsymbol{y} \in \mathcal{D} : 
    \boldsymbol{y} \geq_{vec} \boldsymbol{\omega} + \boldsymbol{\beta} \boldsymbol{x} \right\},
\end{equation*}

\noindent where $\geq_{vec}$ denotes component-wise inequality.
We proceed deriving the reachable set in two-steps, $G_{x,2}$ which includes the points $\boldsymbol{z}=(z_1,z_2)^T$ that can be reached by the process $\{\boldsymbol{v}_t\}$ in two-steps starting from $\boldsymbol{x}$, i.e.,
\begin{equation*}
    G_{x,2} = \left\{ \boldsymbol{z} \in \mathcal{D}: \boldsymbol{z} \geq_{vec} \boldsymbol{\omega} + \boldsymbol{\beta} \boldsymbol{y}, \quad  \boldsymbol{y} \in G_{x,1} \right\}.
\end{equation*}

Similarly, the reachable set in $k$-steps starting from $\boldsymbol{x}$ can be recursively obtained as follows  

\begin{equation*}
    G_{x,k} = \left\{ \boldsymbol{z} \in \mathcal{D}: \boldsymbol{z} \geq_{vec} \boldsymbol{\omega} + \boldsymbol{\beta} \boldsymbol{y}, \quad  \boldsymbol{y} \in G_{x,k-1} \right\}.
\end{equation*}

By induction and using Proposition 1.5.31 \cite{hubbard2009vector} we can derive a more compact expression for $G_{x,k}$, that is: 

\begin{equation} \label{eq_gxk}
    G_{x,k} = \left\{ \boldsymbol{z} \in \mathcal{D} : \boldsymbol{z} \geq_{vec}  \left(\boldsymbol{I}_2 - \boldsymbol{\beta}^k \right) \left(\boldsymbol{I}_2 - \boldsymbol{\beta} \right)^{-1} \boldsymbol{\omega} + \boldsymbol{\beta}^k \boldsymbol{x}  \right\},
\end{equation}

\noindent where $\boldsymbol{I}_2$ denotes a $2\times2$ identity matrix and $(\cdot)^{-1}$ denotes matrix inversion. 

We can also define the limit set $\Omega$ that can be reached with an infinite number of steps as 

\begin{equation} \label{eq_omegone}
    \Omega = \left\{ \boldsymbol{z} \in \mathcal{D} : \boldsymbol{z} \geq_{vec} (\boldsymbol{I}_2 - \boldsymbol{\beta})^{-1} \boldsymbol{\omega}  \right\}.
\end{equation}

We note that the set $\Omega$ does not depend on $\boldsymbol{x}$. Now that we have derived the reachable sets in any step, we proceed studying the one-step transition probability. Given an initial value $\boldsymbol{x} \in \mathcal{D}$ and  $\boldsymbol{y} \in G_{x,1}$, the one-step transition probability is given by

\begin{equation} \label{eq_tran_prob}
     P\left(\boldsymbol{v}_1 \leq \boldsymbol{y}\middle | \boldsymbol{v}_0=\boldsymbol{x} \right) 
     = P \left( \boldsymbol{\omega} + 
    \boldsymbol{\beta} \boldsymbol{v}_0
    + \boldsymbol{\alpha} \boldsymbol{A}_1 
     \leq \boldsymbol{y} \middle| \boldsymbol{v}_0 = \boldsymbol{x} \right), 
\end{equation}

\noindent where $\boldsymbol{A}_1 = (A_{1,1},A_{2,1})^T $ with $ A_{1,1} = \left(W_{1,1}-\gamma_1 \sqrt{v_{1,0}} \right)^2$ and $ A_{2,1} =  \left(W_{2,1}-\gamma_2 \sqrt{v_{2,0}} \right)^2$. 

Using the independence between $A_{1,1}$ and $A_{2,1}$, we can compute the transition probability density function of $\boldsymbol{A}_t$ given $\boldsymbol{v}_0$ as follows

\begin{equation*}
    p_{\boldsymbol{A} | \boldsymbol{v}_0}( a_1,a_2 |  \boldsymbol{v}_0 =  \boldsymbol{x} )=
    \prod_{i=1}^2
    {\frac {1}{\sqrt {2\pi a_i }}}e^{-(a_i+\kappa_i )/2}\cosh({\sqrt {\kappa_i a_i}}) \quad a_i > 0,
\end{equation*}

\noindent which is the product of two independent non-central chi-squared probability density functions with one degree of freedom and non-centrality parameters $\kappa_1 = \gamma^2_1 x_{1} $ and $\kappa_2 = \gamma^2_2 x_{2} $.

Using the linear transformation $\boldsymbol{M}_t = \boldsymbol{\alpha} \boldsymbol{A}_t$ and based on \eqref{eq_tran_prob} we have

\begin{equation*}
     P\left(\boldsymbol{v}_1 \leq \boldsymbol{y}\middle | \boldsymbol{v}_0=\boldsymbol{x} \right) = P(\boldsymbol{M}_1 \leq \boldsymbol{y} - \boldsymbol{\omega} - \boldsymbol{\beta} \boldsymbol{v}_0  | \boldsymbol{v}_0=\boldsymbol{x} ) = \iint \limits _{D_2 \times D_1}  
     p_{\boldsymbol{M}|\boldsymbol{v}_0} (\boldsymbol{m} | \boldsymbol{v}_0 = \boldsymbol{x}) \text{d}\boldsymbol{m},
\end{equation*}

\noindent where $D_1 = (0, y_1 - \omega_1 - \beta_{11} x_1 - \beta_{12} x_2]$, $D_2 = (0, y_2 - \omega_2 - \beta_{21} x_1 - \beta_{22} x_2]$ and $p_{\boldsymbol{M}|\boldsymbol{v}_0}$ is the one-step transition probability density function of $\boldsymbol{M}_t$ given a starting value $\boldsymbol{v}_0$, which is computed as follows

\begin{equation} \label{eq_trans_density}
     p_{\boldsymbol{M}|\boldsymbol{v}_0} \left(\boldsymbol{m} \middle |\boldsymbol{v}_0 = \boldsymbol{x} \right) =  p_{\boldsymbol{A}|\boldsymbol{v}_0} \left(\boldsymbol{\alpha}^{-1}\boldsymbol{m} \middle | \boldsymbol{v}_0  = \boldsymbol{x} \right) 
     \cdot \abs{ \text{det} (\boldsymbol{\alpha}^{-1}) }, 
\end{equation}

\noindent where the Jacobian determinant $\text{det} (\boldsymbol{\alpha}^{-1}) $ is not null by assumption.

According to Definition \ref{def_2}, it remains to prove that for any set $A$ that satisfies $\psi(A) > 0$ there exists $k \in \mathbb{N}$ such that $P^k(\boldsymbol{x}, A) >0$ for all $\boldsymbol{x} \in \mathcal{D}$. Since $\psi(A) > 0$ the intersection between $A$ and $\Omega$ has positive Lebesgue measure. Moreover, since $\Omega$ is the limit of the $G_{x,k}$ sequence of sets, there exists a set $A^* \subset A$, with $\psi(A^*) >0$, which is included in $G_{x,k}$ for every $\boldsymbol{x}$ and for $k$ sufficiently large. Moreover, for each integer $k \geq 2 $, given an initial state $\boldsymbol{x}$, the $k$-step transition probability density function of reaching state $\boldsymbol{z}$ is given by the Chapman-Kolmogorov equation (\cite{tweedie93}, Theorem 3.4.2) 

\begin{equation*}
p^k_{ \boldsymbol{M} | \boldsymbol{v}_0}( \boldsymbol{m} | \boldsymbol{v}_0 = \boldsymbol{x}) = \iint \limits_{G_{x,k-1}} p^{k-1}_{\boldsymbol{M}|\boldsymbol{v}_0}(\boldsymbol{y} | \boldsymbol{v}_0 = \boldsymbol{x}) \cdot p_{\boldsymbol{M}|\boldsymbol{v}_1}( \boldsymbol{m} | \boldsymbol{v}_1 =  \boldsymbol{y}) d\boldsymbol{y},
\end{equation*}

\noindent where $\boldsymbol{m} = \boldsymbol{z} -\boldsymbol{\omega} - \boldsymbol{\beta}\boldsymbol{v}_1$ for $\boldsymbol{z} \in G_{x,k}$ and $p_{\boldsymbol{M}|\boldsymbol{v}_1}( \boldsymbol{m} | \boldsymbol{v}_1 =  \boldsymbol{y})$ denotes the transition probability density function of $\boldsymbol{M}_t$ given $\boldsymbol{v}_1$. By mathematical induction, we deduce that $p^k_{\boldsymbol{M}|\boldsymbol{v}_0}(\boldsymbol{m} | \boldsymbol{v}_0 = \boldsymbol{x})$ is positive for every $\boldsymbol{z} \in A^*$ so that $P^k(\boldsymbol{x},A^*) > 0$ and $P^k(\boldsymbol{x},A) > 0$. 

Having proved $\psi$-irreducibility, we now proceed to study the aperiodicity of the Markov chain $\{ \boldsymbol{v}_t \} $. The aperiodicity is proved if $d=1$ according to Definition \ref{def_aperiod}. By contradiction, if we assume a period $d \geq 2$, 
then there exists a $d$-cycle of $\psi$-positive disjoint sets $E_0, \ldots, E_{d-1}$ such that $P(\boldsymbol{x}, E_{i+1}) = 1$ for every $\boldsymbol{x} \in E_i$ with $i=0,\ldots,d-1$ and such that $V = \left[\bigcup_{i = 0}^{d-1} E_i \right]^c$ has null $\psi$ measure. This implies that when starting our chain $\{ \boldsymbol{v}_t \} $ at any initial state $\boldsymbol{x} \in E_0$ we have $P(\boldsymbol{x}, E_1) = 1$ so that $\psi(E_1 \cap G_{x,1}) >0$ and that $\psi(E_i \cap G_{x,1}) = 0$ for every $i \neq 1$. Then, similarly at a second step, we would have $P(\boldsymbol{y}, E_2) = 1$ for $\boldsymbol{y} \in E_1$ implying $\psi(E_2 \cap G_{x,2}) >0$ and  $\psi(E_i \cap G_{x,2}) = 0$ for every $i \neq 2$ and for initial states $\boldsymbol{x} \in E_0$ such that our chain is in $E_1$ after one step.  But in our case, we know that the intersection $G_{x,1} \cap G_{x,2}$ has non-null $\psi$ measure since $G_{x,1}$ and $G_{x,2}$ are unbounded sets of the type $[a,\infty) \times [b,\infty)$ for some $a,b \in \mathbb{R}_+$ with non-empty intersection with $\Omega$. Therefore, $\psi(G_{x,1} \cap G_{x,2})>0$. Furthermore, 

\begin{equation*}
    \psi(G_{x,1} \cap G_{x,2}) = \psi \left( (G_{x,1} \cap G_{x,2}) \cap \left( \bigcup_{i = 0}^{d-1} E_i \cup V \right) \right) = \sum_{i=0}^{d-1} \psi(G_{x,1} \cap G_{x,2} \cap E_i) >0.
\end{equation*}
    
    Hence, there exists an integer $i=0,\ldots,d-1$ such that $\psi(G_{x,1} \cap G_{x,2} \cap E_i)>0$. This contradicts the fact that $\psi(E_i \cap G_{x,1}) = 0$ for every $i \neq 1$ and $\psi(E_i \cap G_{x,2}) = 0$ for every $i \neq 2$.
$\hfill \square$

\textit{Proof of Proposition \ref{prop_small}.} First, let us define 
\begin{equation} \label{eq_rbarrato}
    \Bar{r} = \max\left({ \sqrt{\omega_1^2 + \omega_2^2} , \underline{\Omega} } \right),
\end{equation}
\noindent where $\underline{\Omega} = \frac{\max \left[ (1-\beta_{22}) \omega_1 + \beta_{12} \omega_2 ,  \quad \beta_{21} \omega_1 + (1-\beta_{11}) \omega_2 \right ]}{ ( (1-\beta_{11}) (1-\beta_{22}) - \beta_{12} \beta_{21} ) }$ represents the the maximum coordinate value of the closest point to the origin of the set $\Omega$, as defined in \eqref{eq_omegone}. It can be easily shown that $\underline{\Omega} > 0$. Indeed, we have 

\begin{equation*}
    (\boldsymbol{I}_2 - \boldsymbol{\beta})^{-1} \boldsymbol{\omega}= \frac{1}{\text{det}(\boldsymbol{I}_2 - \boldsymbol{\beta})} \begin{bmatrix}
        1-\beta_{22} & \beta_{12} \\
        \beta_{21} & 1-\beta_{11}
    \end{bmatrix} \boldsymbol{\omega},
\end{equation*}

\noindent where 

\begin{align*}
    \text{det}(\boldsymbol{I}_2 - \boldsymbol{\beta}) &= 1 - \text{trace}(\boldsymbol{\beta}) + \text{det}(\boldsymbol{\beta}) \\
    &= 1 - (\varlambda_1 + \varlambda_2) + \varlambda_1 \varlambda_2 \\
    &= (1 - \varlambda_1) ( 1 - \varlambda_2) > 0,
\end{align*}

\noindent with $\varlambda_{1}$ and $\varlambda_2$ denoting the eigenvalues of $\boldsymbol{\beta}$, which are guaranteed strictly smaller than one in absolute value by $\rho(\boldsymbol{\beta})<1$ .

For proving that the compact set $C= \{ \boldsymbol{x} \in \mathcal{D} : \norm{\boldsymbol{x}} \leq r \}$, for any $r > \Bar{r}$ , is small we use Proposition 2.11 of \cite{nummelin84}. We note that by letting $r > \Bar{r}$ we ensure $C$ to be non-empty and $C \in \mathcal{B}^+$, given that $C$ has $\mu_{Leb}$-positive intersection with $\Omega$. The proposition states that a set $C \in \mathcal{B}^+$ is small, if there is a set $B \in \mathcal{B}^+$, such that for all $A \in \mathcal{B}^+$ with $A \subseteq B$,

    \begin{equation*}
    \inf_{\boldsymbol{x} \in C} \sum_{n=0}^L P^n (\boldsymbol{x}, A) > 0,
    \end{equation*}
    
\noindent for some non-negative integer $L$.
    
We prove that the condition above holds for some $L>1$. The first step is to find the set $B$ with $\psi(B) > 0$ which implies that $B \in \mathcal{B}^+$ as $\psi$ is the irreducibility measure.  Let us define $B = (d_1, \infty) \times (d_2, \infty)$ with $(d_1, d_2)^T =  (\boldsymbol{I}_2 - \boldsymbol{\beta})^{-1} \boldsymbol{\omega} + \boldsymbol{\beta} \cdot (r,r)^T$ so that $B \subseteq G_{x,k}$ for each $\boldsymbol{x} \in C$ and $k>0$. 

By the $\psi$-irreducibility proved in Lemma \ref{lemma1}, for $\boldsymbol{x} \in C$, we have a positive probability of visiting all sets $A \subseteq B$ with $\psi(A) > 0$, {i.e.}, $P^k(\boldsymbol{x},A) > 0$ for some $k>0$. Lastly, the proof is concluded since we know that a positive and real-valued continuous function on $C$ has a positive infimum on $C$.
$\hfill\square$

\textit{Proof of Lemma \ref{ineq_a}.} The result is built upon Lemma 1 in \cite{hernandez19}, which states that

\begin{equation*} 
    \left( W - \gamma \sqrt{v} \right)^2 \leq \left( \gamma + W^2 \mathds{1}_{(\gamma W < 0)} \right) v  + \left(  \abs{\gamma} + \sqrt{ \gamma^2 + W^2 } \right)^2, \quad v >0, W \in \mathds{R}.
\end{equation*}

Using Assumption 1, the above inequality can be generalized as follows

\begin{align} \label{eq:hern1}
     \alpha_{11} \left( W_{1,t} - \gamma_1 \sqrt{v_{1,t}} \right)^2 &+ \alpha_{12} \left( W_{2,t} - \gamma_2 \sqrt{v_{2,t}} \right)^2   \\ \notag
      \leq & \alpha_{11} \left[ \left( \gamma_1 + W^2_{1,t} \mathds{1}_{(\gamma_1 W_{1,t} < 0)} \right) v_{1,t}  +\left(  \abs{\gamma_1} + \sqrt{ \gamma_1^2 + W_{1,t}^2 } \right)^2 \right]  \\ \notag
     +&\alpha_{12}  \left[ \left( \gamma_2 + W_{2,t}^2 \mathds{1}_{(\gamma_2 W_{2,t} < 0)} \right) v_{2,t}  +  \left(  \abs{\gamma_2} + \sqrt{ \gamma_2^2 + W_{2,t}^2 } \right)^2 \right],
\end{align}

\noindent where $v_{1,t}, v_{2,t} \in \mathcal{D}$ and $  W_{1,t},  W_{2,t} \in \mathds{R}$. Similarly, we have:

\begin{align} \label{eq:hern2}
     \alpha_{21} \left( W_{1,t} - \gamma_1 \sqrt{v_{1,t}} \right)^2 &+  \alpha_{22} \left( W_{2,t} - \gamma_2 \sqrt{v_{2,t}} \right)^2   \\ \notag
        \leq & \alpha_{21}  \left[ \left( \gamma_1 + W^2_{1,t} \mathds{1}_{(\gamma_1 W_{1,t} < 0)} \right) v_{1,t}  +  \left(  \abs{\gamma_1} + \sqrt{ \gamma_1^2 + W_{1,t}^2 } \right)^2 \right] 
        \\ \notag
      + &\alpha_{22} \left[ \left( \gamma_2 + W_{2,t}^2 \mathds{1}_{(\gamma_2 W_{2,t} < 0)} \right) v_{2,t}  +  \left(  \abs{\gamma_2} + \sqrt{ \gamma_2^2 + W_{2,t}^2 } \right)^2 \right].
\end{align}

Using equations \eqref{eq:hern1} and \eqref{eq:hern2} we can derive the following result:

\begin{align*}
    \norm{\boldsymbol{v}_t} & \leq 
    \norm{
    \boldsymbol{\omega} + \left\{ \boldsymbol{B} +  \boldsymbol{b}(\boldsymbol{W}_t) \right \} \boldsymbol{v}_{t-1} + \boldsymbol{\alpha }
    \boldsymbol{f}(\boldsymbol{W}_t) },
\end{align*}

\noindent where $\boldsymbol{B}$ is specified in \eqref{eq_auto}.
$\hfill \square$

\textit{Proof of Theorem \ref{theorem}.} We first show that the so-called drift condition is satisfied with the function $Q(\boldsymbol{v})=\norm{\boldsymbol{v}}$. Then, we can extend geometric ergodicity property of the variance process $\{\boldsymbol{v}_t\}$ to the return process thanks to \cite{carrasco02}. 

For verifying the drift condition we use the $m$-step criterion introduced by \cite{tjostheim90}. Specifically, we consider $m=1$ and we verify Conditions (3.7) and (3.8) in \cite{tjostheim90} using a small set $C= \{ \boldsymbol{x} \in \mathcal{D} : \norm{\boldsymbol{x}} \leq r \}$ for $r > \Bar{r}$ sufficiently large with complement $C^c$ and a non-negative test function $Q$:

\begin{align} 
    \mathbb{E} \left[ Q( \boldsymbol{v}_t)\mathds{1}_{(Q( \boldsymbol{v}_t) \in C^c)}
    \middle | 
     \boldsymbol{v}_{t-1} = \boldsymbol{x} \right] \leq M, \quad \boldsymbol{x} \in C, \label{eq_3.7} \\
     R \cdot \mathbb{E} \left[ Q( \boldsymbol{v}_t) 
    \middle | 
     \boldsymbol{v}_{t-1} = \boldsymbol{x} \right] \leq Q(\boldsymbol{x}) - \epsilon, \quad \boldsymbol{x} \in C^c, \label{eq_3.8}
\end{align}

\noindent for some $M < \infty$, $\epsilon >0$ and some $R>1$. 

Using the independence between $\boldsymbol{b}(\boldsymbol{W}_t)$ and $\boldsymbol{f}(\boldsymbol{W}_s)$ for $s<t$, together with Lemma \ref{ineq_a}, we obtain:

\begin{align} \notag
     \mathbb{E} \left[ Q( \boldsymbol{v}_t) \middle | 
     \boldsymbol{v}_{t-1} = \boldsymbol{x} \right]  &\leq  \mathbb{E} \left[ \norm { \boldsymbol{\omega} + \left\{ \boldsymbol{B} + \boldsymbol{b} (\boldsymbol{W}_t) \right \} \boldsymbol{v}_{t-1} + \boldsymbol{\alpha}\boldsymbol{f}(\boldsymbol{W}_t)
    } \middle | \boldsymbol{v}_{t-1} = \boldsymbol{x} \right] 
    \\ 
    &\leq  \varphi Q(\boldsymbol{x}) + c, \label{eq_drift}
 \end{align}

\noindent where $c=\left( \norm{\boldsymbol{\omega}} + \mathbb{E} \left[ \norm{ \boldsymbol{\alpha}\boldsymbol{f}(\boldsymbol{W}_t)}\right]  \right) < \infty$  since $ \mathbb{E} \left[ \norm{ \boldsymbol{\alpha}\boldsymbol{f}(\boldsymbol{W}_t)}\right]  < \infty$ and the second inequality follows from the subadditivity and submultiplicative properties of the Euclidean norm. 

Using inequality \eqref{eq_drift}, we can easily verify condition 
\eqref{eq_3.7}, since $\mathbb{E} \left[ Q( \boldsymbol{v}_t) \middle | 
     \boldsymbol{v}_{t-1} = \boldsymbol{x} \right] \leq M$, for $\boldsymbol{x} \in C$ and  $Q$ is nonnegative. Let us prove inequality \eqref{eq_3.8}. We can find a constant $R>1$, such that $R \varphi < 1$, so that after multiplying inequality \eqref{eq_drift} by $R$ and by adding and substracting $\norm{\boldsymbol{x}}$ we obtain

\begin{equation*}
    R \cdot \mathbb{E} \left[ Q( \boldsymbol{v}_t) \middle | \boldsymbol{v}_{t-1} = \boldsymbol{x} \right] \leq \norm{\boldsymbol{x}} + \left(R \varphi - 1 \right) \norm{\boldsymbol{x}} + Rc , \quad \boldsymbol{x} \in C^c ,
\end{equation*}

\noindent which satisfies \eqref{eq_3.8} for $r > \Bar{r}$ large enough. 

%We note that as remarked in \cite{tjostheim90}, condition \eqref{eq_3.8} is stronger than condition (3.6) of \cite{tjostheim90}, which is needed for $\{ \boldsymbol{v}_t \}$ to be positive Harris recurrent. Using aperiodicity and positive Harris recurrent $\{ \boldsymbol{v}_t \}$ we obtain (Harris) ergodicity.

Now that we have proved the conditions \eqref{eq_3.7}-\eqref{eq_3.8}, according to \cite{tjostheim90} we can state that $\{\boldsymbol{v}_t\}$ is geometrically ergodic and if it is initialized from its invariant measure $\pi$, then it is strictly stationary. Finally, the geometric ergodicity can be extended to the Markov chain $\{(R_t, \boldsymbol{v}_t)\}$ by  Proposition 4(i) of \cite{carrasco02}, so that $\{R_t\}$ is strictly stationary and geometrically ergodic. 
$\hfill \square$

\textit{Proof of Proposition \ref{prop_cont}.} Following the discretization scheme of \cite{nelson90}, we consider the stochastic difference equations \eqref{eq:1} and \eqref{eq:2} as we partition time more and more finely. We allow the parameters of the system $\boldsymbol{\omega},\boldsymbol{\beta}, \boldsymbol{\alpha}$ and $\boldsymbol{\gamma}$ to depend on the time partition $h>0$ and make both the drift term of equation \eqref{eq:1} and the variance of $Z_{1,t}$ and $Z_{2,t}$ proportional to $h$, so that

\begin{equation*} 
    {_h}R_{kh} =  r_h + \lambda_{h} ( {_h}v_{1,kh} +  {_h}v_{2,kh} ) + \sqrt{{_h}v_{1,kh}} \cdot {_h}Z_{1,kh} + \sqrt{{_h}v_{2,kh}} \cdot {_h}Z_{2,kh}, 
\end{equation*}

\noindent where ${_h}v_{1,kh}$ and $ {_h}v_{2,kh}$ denote a sequence of discrete times that depend on both $h$ and the (discrete) time index $kh$, $k = 0,1,2,\ldots$ and ${_h}Z_{i,kh} \overset{i.i.d.}{\sim} N(0,h)$, $i = 1,2$, with ${_h}Z_{1,kh} \perp {_h}Z_{2,kh}$. After computing the square and using matrix notation, equation \eqref{eq:2} is rewritten as follows

\begin{equation*}
    {_h}\boldsymbol{v}_{kh} = \boldsymbol{\omega}_h + (\boldsymbol{\beta}_h + \boldsymbol{\alpha}_h \boldsymbol{\gamma}^T_h \boldsymbol{\gamma}_h) {_h}\boldsymbol{v}_{(k-1)h} + h^{-1/2}\boldsymbol{\alpha}_h  \begin{bmatrix}
       {_h}Z^2_{1,(k-1)h}  \\
       {_h}Z^2_{2,(k-1)h} 
    \end{bmatrix} - 2 h^{-1} \boldsymbol{\alpha}_h \boldsymbol{\gamma}_h \begin{bmatrix}
         {_h}Z_{1,(k-1)h} \sqrt{{_h}v_{1,(k-1)h}} \\
         {_h}Z_{2,(k-1)h} \sqrt{{_h}v_{2,(k-1)h}}
    \end{bmatrix},
\end{equation*}

\noindent where ${_h}\boldsymbol{v}_{kh} = \begin{bmatrix}
    {_h}v_{1,kh} & {_h}v_{2,kh}
\end{bmatrix}^T$, 
$\boldsymbol{\omega}_h = \begin{bmatrix}
    \omega_{1,h} & \omega_{2,h}
\end{bmatrix}^T$, 

\begin{equation*}
    \boldsymbol{\beta}_h = \begin{bmatrix}
        \beta_{11,h}  & \beta_{12,h}  \\
        \beta_{21,h}  & \beta_{22,h} 
    \end{bmatrix}, \quad
    \boldsymbol{\alpha}_h=
    \begin{bmatrix}
    \alpha_{11,h} & 
    \alpha_{12,h} \\
    \alpha_{21,h}  & 
    \alpha_{22,h} \\
    \end{bmatrix}, \quad
    \boldsymbol{\gamma}_h
    \begin{bmatrix}
    \gamma_{1,h} & 0 \\
    0 & \gamma_{2,h}
    \end{bmatrix},
\end{equation*}

\noindent where, for all $h$, Assumption 1 ensures that the process $ {_h}\boldsymbol{v}_{kh}$ remains positive with probability one. 

We define continuous time processes $\{ {_h}R_t \}$ and $\{{_h}\boldsymbol{v}_t\}$ as

\begin{equation*}
    {_h}R_t = {_h}R_{kh}, \quad {_h}\boldsymbol{v}_t = {_h}\boldsymbol{v}_{kh}, \quad kh \leq t < (k+1)h.
\end{equation*}

We derive the drift per unit of time, conditioned on information at time $(k - 1)h$, for the return as

\begin{equation} \label{eq_first_mom1}
    \mathbb{E}_{kh}\left[ h^{-1} {_h}R_{kh} \right] = r_h + \lambda_{h} ( {_h}v_{1,kh} +  {_h}v_{2,kh}),  
\end{equation}

\noindent and for the variance process as

\begin{equation}\label{eq_first_mom2}
    \mathbb{E}_{kh}\left[ h^{-1} ( {_h}\boldsymbol{v}_{(k+1)h} - {_h}\boldsymbol{v}_{kh} ) \right] =    h^{-1}  \boldsymbol{\omega}_h +  h^{-1}
   (\boldsymbol{\beta}_h + \boldsymbol{\alpha}_h \boldsymbol{\gamma}^T_h \boldsymbol{\gamma}_h - \boldsymbol{I}_2)
    {_h}\boldsymbol{v}_{k h}  + 
    h^{-1/2}
    \begin{bmatrix}
    \alpha_{11,h} + 
    \alpha_{12,h} \\
    \alpha_{21,h} + 
    \alpha_{22,h} \\
    \end{bmatrix},
\end{equation}

\noindent where $\mathbb{E}_{kh} [\cdot] = \mathbb{E}[ \cdot | M_{kh}]$ and $M_{kh}$ denotes the $\sigma-$algebra generated by  ${_h}R_0, \ldots, {_h}R_{(k-1)h}$ and with ${_h}\boldsymbol{v}_0, \ldots, {_h}\boldsymbol{v}_{kh}$. For the drift per unit of time to converge the following limits must exist and be finite

\begin{align}\label{eq_limits}
    &\lim_{h \rightarrow 0} h^{-1} r_h = r, \quad \lim_{h \rightarrow 0} h^{-1} \lambda_{i,h} = \lambda_i, \quad  \lim_{h \rightarrow 0} h^{-1} \boldsymbol{\omega}_h = \boldsymbol{\omega},  \\ \label{eq_limits3}
    &\lim_{h \rightarrow 0} h^{-1/2} 
    \begin{bmatrix}
        \alpha_{11,h} + \alpha_{12,h} \\
        \alpha_{21,h} + \alpha_{22,h}
    \end{bmatrix} = \begin{bmatrix}
        \alpha_{11} + \alpha_{12} \\
        \alpha_{21} + \alpha_{22}
    \end{bmatrix}, \quad \lim_{h \rightarrow 0} h^{-1} (  \boldsymbol{I}_2 - \boldsymbol{\beta}_h - \boldsymbol{\alpha}_h \boldsymbol{\gamma}^T_h \boldsymbol{\gamma}_h ) = \boldsymbol{\theta}.
\end{align}

Similarly, by letting ${_h}\mu_{kh} = r_h + \lambda_{h} ( {_h}v_{1,kh} + {_h}v_{2,kh} )$ for convenience, we can derive the second conditional moment per unit of time for the return:

\begin{align}\label{eq_sec_mom1}
    \mathbb{E}_{kh} \left[ h^{-1} R^2_{kh}  \right] = {_h}\mu^2_{kh} + {_h}v_{1,kh} + {_h}v_{2,kh},
\end{align}

\noindent the conditional covariance per unit of time:

\begin{align}\label{eq_sec_mom2}
    \mathbb{E}_{kh}\left[ h^{-1} {_h}R_{kh} ( {_h}\boldsymbol{v}_{(k+1)h} - {_h}\boldsymbol{v}_{kh} ) \right] &=  h^{-1} \mathbb{E}_{kh} \bigg{[} ( {_h}\mu_{kh} + {_h}Z_{1,kh}\sqrt{{_h}v_{1,kh}} + {_h}Z_{2,kh} \sqrt{{_h}v_{2,kh}})  \\ \notag
    &\times \bigg{(} \boldsymbol{\omega}_h + (\boldsymbol{\beta}_h + \boldsymbol{\alpha}_h \boldsymbol{\gamma}^T_h \boldsymbol{\gamma}_h - \boldsymbol{I}_2 ) {_h}\boldsymbol{v}_{kh}
    + h^{-1/2} \boldsymbol{\alpha}_h \begin{bmatrix}
        {_h}Z^2_{1,kh} \\
        {_h}Z^2_{2,kh}
    \end{bmatrix} \\
    &- 2 h^{-1} \boldsymbol{\alpha}_h \boldsymbol{\gamma}_h 
    \begin{bmatrix}
        {_h}Z_{1,kh}\sqrt{{_h}v_{1,kh}} \\ \notag
        {_h}Z_{2,kh}\sqrt{{_h}v_{2,kh}}
    \end{bmatrix} 
    \bigg{)} \bigg{]} \\\notag
    &=
    \boldsymbol{\omega}_h \cdot {_h}\mu_{kh} \cdot {_h}\boldsymbol{v}_{kh} + (\boldsymbol{\beta}_h + \boldsymbol{\alpha}_h \boldsymbol{\gamma}^T_h \boldsymbol{\gamma}_h - \boldsymbol{I}_2 ) {_h}\boldsymbol{v}_{kh} {_h}\mu_{kh}  \\ \notag
    &+ h^{1/2} {_h}\mu_{kh}  \begin{bmatrix}
        \alpha_{11,h} + \alpha_{12,h} \\
        \alpha_{21,h}
        + \alpha_{22,h}
    \end{bmatrix} - 2 h^{-1} \boldsymbol{\alpha}_h \boldsymbol{\gamma}_h {_h}\boldsymbol{v}_{kh},  
\end{align}

\noindent and the second conditional moment of the variance per unit of time: 

\begin{align} \label{eq_sec_mom3}
     &\mathbb{E}_{kh} \left[ h^{-1} ( {_h}\boldsymbol{v}_{(k+1)h} - {_h}\boldsymbol{v}_{kh} ) ( {_h}\boldsymbol{v}_{(k+1)h} - {_h}\boldsymbol{v}_{kh} )^T \right] = \\ \notag
     &= h^{-1} \mathbb{E}_{kh} \Bigg[ 4 h^{-2} \left( \boldsymbol{\alpha}_h \boldsymbol{\gamma}_h \begin{bmatrix}
         {_h}Z_{1,kh} \sqrt{ {_h}v_{1,kh} } \\
         {_h}Z_{2,kh} \sqrt{ {_h}v_{2,kh} } 
     \end{bmatrix} \right) \left( \boldsymbol{\alpha}_h \boldsymbol{\gamma}_h \begin{bmatrix}
         {_h}Z_{1,kh} \sqrt{{_h}v_{1,kh}} \\ 
         {_h}Z_{2,kh} \sqrt{{_h}v_{2,kh}} 
     \end{bmatrix} \right)^T    \\ \notag
     &+ \left( \boldsymbol{\omega}_h +  (\boldsymbol{\beta}_h + \boldsymbol{\alpha}_h \boldsymbol{\gamma}^T_h \boldsymbol{\gamma}_h - \boldsymbol{I}_2 ) {_h}\boldsymbol{v}_{kh} + \boldsymbol{\alpha}_h h^{-1/2} \begin{bmatrix}
         {_h}Z^2_{1,kh}  \\ 
         {_h}Z^2_{2,kh} 
     \end{bmatrix}  \right )  \\ \notag
     &\times \left( \boldsymbol{\omega}_h +  (\boldsymbol{\beta}_h + \boldsymbol{\alpha}_h \boldsymbol{\gamma}^T_h \boldsymbol{\gamma}_h - \boldsymbol{I}_2 ) {_h}\boldsymbol{v}_{kh} + \boldsymbol{\alpha}_h h^{-1/2}\begin{bmatrix}
         {_h}Z^2_{1,kh}  \\ 
         {_h}Z^2_{2,kh} 
     \end{bmatrix} \right )^T +  \boldsymbol{\xi}_0 \Bigg]
       \\ \notag
     &=  4h^{-2} \begin{bmatrix}
         \alpha^2_{11,h} \gamma^2_{1,h} {_h}v_{1,kh} + \alpha^2_{12,h} \gamma^2_{2,h} {_h}v_{2,kh} & 
         \alpha_{11,h}
         \alpha_{21,h}
         \gamma^2_{1,h} {_h}v_{1,kh} + \alpha_{12,h}
         \alpha_{22,h}
         \gamma^2_{2,h} 
         {_h}v_{2,kh} \\
         \alpha_{11,h}
         \alpha_{21,h}
         \gamma^2_{1,h} {_h}v_{1,kh} + \alpha_{12,h}
         \alpha_{22,h}
         \gamma^2_{2,h} 
         {_h}v_{2,kh} & \alpha^2_{21,h} \gamma^2_{1,h} {_h}v_{1,kh} + \alpha^2_{22,h}  \gamma^2_{2,h} {_h}v_{2,kh}
     \end{bmatrix} \\ \notag 
     \\ &+ h^{-1} \left( \boldsymbol{\omega}_h +  (\boldsymbol{\beta}_h + \boldsymbol{\alpha}_h \boldsymbol{\gamma}^T_h \boldsymbol{\gamma}_h - \boldsymbol{I}_2 ) {_h}\boldsymbol{v}_{kh} + 3 h^{1/2} \begin{bmatrix}
         \alpha_{11,h} + \alpha_{12,h}  \\ 
         \alpha_{21,h} + \alpha_{22,h} 
     \end{bmatrix}  \right )  \notag \\ \notag
     &\times \left( \boldsymbol{\omega}_h +  (\boldsymbol{\beta}_h + \boldsymbol{\alpha}_h \boldsymbol{\gamma}^T_h \boldsymbol{\gamma}_h - \boldsymbol{I}_2 ) {_h}\boldsymbol{v}_{kh} + 3 h^{1/2} \begin{bmatrix}
         \alpha_{11,h} + \alpha_{12,h}  \\ 
         \alpha_{21,h} + \alpha_{22,h} 
     \end{bmatrix} \right )^T, 
\end{align}

\noindent where $\boldsymbol{\xi}_0$ is defined as a $(2 \times 1)$ vector containing terms of equation \eqref{eq_sec_mom3} that satisfy $ \mathbb{E}_{kh} \left[ \boldsymbol{\xi}_0 \right] = \boldsymbol{0}$. 

Assuming that 

\begin{equation}\label{eq_limits2}
    \lim_{h \rightarrow 0} h^{-1/2} \boldsymbol{\gamma}_h = \boldsymbol{ \gamma},
\end{equation}

\noindent and substituting \eqref{eq_limits}-\eqref{eq_limits3} into equations \eqref{eq_sec_mom1}-\eqref{eq_sec_mom3} we obtain

\begin{align}
    &\mathbb{E}_{kh} \left[ h^{-1} {_h}R^2_{kh}  \right] = {_h}v_{1,kh} + {_h}v_{2,kh} + o(1), 
    \label{eq_sec_mom1_l}\\[6pt]
    &\mathbb{E}_{kh}\left[ h^{-1} {_h}R_{kh} ( {_h}\boldsymbol{v}_{(k+1)h} - {_h}\boldsymbol{v}_{kh} ) \right] = - 2  \boldsymbol{\alpha} \boldsymbol{\gamma} \cdot {_h}\boldsymbol{v}_{kh} + \boldsymbol{o}(1), \label{eq_sec_mom2_l} \\[6pt] 
    &\mathbb{E}_{kh} \left[ h^{-1} ( {_h}\boldsymbol{v}_{(k+1)h} - {_h}\boldsymbol{v}_{kh} ) ( {_h}\boldsymbol{v}_{(k+1)h} - {_h}\boldsymbol{v}_{kh} )^T \right] = \label{eq_sec_mom3_l}
    \\ \notag
    &4 \begin{bmatrix}
         \alpha^2_{11} \gamma^2_1 {_h}v_{1,kh} + \alpha^2_{12} \gamma^2_2 {_h}v_{2,kh} & 
         c_{kh} \\
         c_{kh} & \alpha^2_{21} \gamma^2_1 {_h}v_{1,kh} + \alpha^2_{22}  \gamma^2_2 {_h}v_{2,kh}
     \end{bmatrix} 
     + \boldsymbol{o}(1), 
\end{align}

\noindent where $c_{kh} = \alpha_{11}
         \alpha_{21}
         \gamma^2_1 {_h}v_{1,kh} + \alpha_{12}
         \alpha_{22}
         \gamma^2_2 
         {_h}v_{2,kh}$. We can easily find $\{r_h , \lambda_h, \boldsymbol{\omega}_h, \boldsymbol{\alpha}_h, \boldsymbol{\gamma}_h , \boldsymbol{\beta}_h\}$ sequences that satisfy \eqref{eq_limits}-\eqref{eq_limits3} and \eqref{eq_limits2}, for instance, $r_h = r \cdot h$, $\lambda_h = \lambda \cdot h$, $\boldsymbol{\omega}_h = \boldsymbol{\omega} \cdot h, \boldsymbol{\alpha}_h = \sqrt{h} \boldsymbol{\alpha}, \boldsymbol{\gamma}_h = \sqrt{h} \boldsymbol{\gamma}, \boldsymbol{\beta}_h = \boldsymbol{I}_2 - {h}^{3/2} \boldsymbol{\alpha} \left(\boldsymbol{\gamma}^T\boldsymbol{\gamma}\right)  - \boldsymbol{\theta}h$. Let us denote the instantaneous drift function, i.e., the right hand side limit for $h$ tending to zero of equations \eqref{eq_first_mom1}-\eqref{eq_first_mom2}, as $b$:
         
\begin{equation} \label{eq_b_cont}
    b(R_t, v_{1,t}, v_{2,t}) 
    =
    \begin{bmatrix} 
        r + \lambda (v_{1,t} + v_{2,t}) \\
        \omega_1 + \alpha_{11} + \alpha_{12} - \theta_{11} v_{1,t} - \theta_{12} v_{2,t} \\
        \omega_2 + \alpha_{21} + \alpha_{22} - \theta_{21} v_{1,t} - \theta_{22} v_{2,t}
    \end{bmatrix},
\end{equation}

\noindent and denote the instantaneous covariance matrix, i.e., as the right hand side limit for $h$ tending to zero of equations \eqref{eq_sec_mom1_l}-\eqref{eq_sec_mom3_l}, as $a$:
 
\begin{align} \label{eq_a_cont}
    &a(R_t, v_{1,t} , v_{2,t}) = \\ \notag
    &\begin{bmatrix}
        v_{1,t} + v_{2,t} & - 2 (\alpha_{11} \gamma_1 v_{1,t} +  \alpha_{12} \gamma_2 v_{2,t}) & - 2 (\alpha_{21} \gamma_1 v_{1,t} +  \alpha_{22} \gamma_2 v_{2,t})  \\
       - 2 (\alpha_{11} \gamma_1 v_{1,t} +  \alpha_{12} \gamma_2 v_{2,t})  & 4 (\alpha^2_{11} \gamma^2_1 v_{1,t} + \alpha^2_{12}  \gamma^2_2 v_{2,t}) & 
        4(\alpha_{11}
         \alpha_{21}
         \gamma^2_1 v_{1,t} + \alpha_{12}
         \alpha_{22}
         \gamma^2_2 
         v_{2,t}) \\
       - 2 (\alpha_{21} \gamma_1 v_{1,t} +  \alpha_{22} \gamma_2 v_{2,t}) & 
        4(\alpha_{11}
         \alpha_{21}
         \gamma^2_1 v_{1,t} + \alpha_{12}
         \alpha_{22}
         \gamma^2_2 
         v_{2,t}) & 4( \alpha^2_{21} \gamma^2_1 v_{1,t} + \alpha^2_{22} \gamma^2_2 v_{2,t})
    \end{bmatrix}.
\end{align}

Using Theorem 2.1 of \cite{nelson90}, the limits in \eqref{eq_limits}-\eqref{eq_limits3} and \eqref{eq_limits2}, and definitions  \eqref{eq_b_cont} and \eqref{eq_a_cont}, we have that the processes ${_h}R_{kh}$ and ${_h}\boldsymbol{v}_{kh}$ converge to continuous time processes $R_t$ and $\boldsymbol{v}_t$, solutions of the following system of stochastic differential equations:

\begin{align*} 
    dR_t &=  (r + \lambda (v_{1,t} + v_{2,t})) dt + \sqrt{v_{1,t}} dW_{1,t} + \sqrt{v_{2,t}} dW_{2, t}, \\ \notag
    dv_{1,t} &= (\omega_1 + \alpha_{11} + \alpha_{12} - \theta_{11} v_{1,t} - \theta_{12} v_{2,t}) dt + 2 \alpha_{11} \gamma_1 \sqrt{v_{1,t}} dW_{3,t} + 2\alpha_{12}  \gamma_2\sqrt{ v_{2,t}} dW_{4,t}, 
    \\ \notag
    dv_{2,t} &= (\omega_2 + \alpha_{21} + \alpha_{22} - \theta_{21} v_{1,t} - \theta_{22} v_{2,t}) dt + 2 \alpha_{21} \gamma_1\sqrt{v_{1,t}} dW_{3,t} + 2 \alpha_{22} \gamma_2\sqrt{ v_{2,t} } dW_{4,t},
\end{align*}

\noindent where the Brownian motions $W_{i,t}$, $i=1,\ldots,4$,  are independent from the initial values $(R_0, v_{1,0}, v_{2,0})$ and from each other, except for

\begin{equation*}
    \mathbb{E} \left[dW_{1,t}dW_{3,t}\right]= - \text{sign}({\gamma_1})dt , \quad \mathbb{E} \left[dW_{2,t}dW_{4,t}\right]= - \text{sign}({\gamma_2})dt.
\end{equation*}
$\hfill \square$

\textit{Proof of Proposition \ref{prop_option1}.} The equivalent martingale measure $\mathbb{Q}$ must satisfy 
    \begin{equation}
     \mathbb{E} \left[e^{R_{t}} \frac{dQ_{t}}{dP_{t}} \middle| \mathcal{F}_{t-1} \right] = e^r.
\end{equation}

From equations \eqref{eq:1} and \eqref{eq_rn_derivative} we obtain

\begin{equation} \label{eq_prev}
     e^{r + \lambda ({v}_{1,{t}} +  {v}_{2,{t}}) } 
     \frac{ \mathbb{E} \left[ {e^{  (\Lambda + 1) (\sqrt{v_{1,t}} Z_{1,t} +  \sqrt{v_{2,t}} Z_{2,t} })  } 
      \middle| \mathcal{F}_{t-1} \right]}{\mathbb{E} \left[ {e^{  \Lambda (  \sqrt{v_{1,t}} Z_{1,t} +  \sqrt{v_{2,t}} ) Z_{2,t} } } 
      \middle| \mathcal{F}_{t-1} \right]}
      = e^r.
\end{equation}

By taking logarithms we can rewrite \eqref{eq_prev} as

\begin{equation} \label{eq_prev2}
    \lambda ({v}_{1,{t}} + {v}_{2,{t}}) + \ln{ \left(  \frac{\mathbb{E} \left[ {e^{  (\Lambda + 1) (\sqrt{v_{1,t}} Z_{1,t} + \sqrt{v_{2,t}} Z_{2,t}) }} 
      \middle| \mathcal{F}_{t-1} \right]}{\mathbb{E} \left[ {e^{  \Lambda (  \sqrt{v_{1,t}} Z_{1,t} +  \sqrt{v_{2,t}} Z_{2,t}) } } 
      \middle| \mathcal{F}_{t-1} \right]}
     \right)}  = 0.
\end{equation}

The conditional expectations in \eqref{eq_prev2} can be easily computed since $Z_{1,{t}}$ and $Z_{2,{t}}$ are normally distributed and independent, we obtain

\begin{align}
    \mathbb{E} \left[ {e^{  \Lambda (  \sqrt{v_{1,t}} Z_{1,t} +  \sqrt{v_{2,t}} Z_{2,t}) } } 
      \middle| \mathcal{F}_{t-1} \right] &= e^{ \frac{1}{2} \Lambda^2 \left( {v}_{1,{t}}  +  {v}_{2,{t}} \right)},  \label{eq_prev3}
      \\
      \label{eq_prev4}
       \mathbb{E} \left[ {e^{  (\Lambda + 1)  ( \sqrt{v_{1,t}} Z_{1,t} +  \sqrt{v_{2,t}} Z_{2,t} ) } } 
      \middle| \mathcal{F}_{t-1} \right] &=  e^{ \frac{1}{2} (\Lambda + 1 )^2 \left(  {v}_{1,{t}} +  {v}_{2,{t}}  \right)}.
\end{align}

\noindent Therefore, by using \eqref{eq_prev3} and \eqref{eq_prev4} we obtain

\begin{equation}
     \lambda ({v}_{1,{t}} +  {v}_{2,{t}}) + \ln{ \left(  \frac{e ^{  \frac{1}{2}(\Lambda + 1 )^2 \left( {v}_{1,{t}} + {v}_{2,{t}} \right)}}{e^{ \frac{1}{2} \Lambda^2 \left({v}_{1,{t}} + {v}_{2,{t}} \right)}}
     \right)}  = 0,
\end{equation}

\noindent which yields

\begin{equation}
    \left(\lambda + \frac{1}{2} \right) ({v}_{1,{t}} +  {v}_{2,{t}}) + 
    \Lambda ({v}_{1,{t}} + {v}_{2,{t}}) = 0.
\end{equation}

By collecting terms we obtain

\begin{equation}
    \left( \lambda + \frac{1}{2} + \Lambda  \right) ({v}_{1,{t}} + {v}_{2,{t}}) = 0,
\end{equation}

\noindent and, by equating the coeﬀicients of  ${v}_{1,{t}}$ and ${v}_{2,{t}}$ to zero we obtain equation \eqref{eq_Lambda}.
$\hfill \square$

\textit{Proof of Proposition \ref{prop_option2}.} First, we denote by $\mathbb{E}^\mathbb{Q}$ the expectation operator under the risk-neutral measure $\mathbb{Q}$, so that the risk-neutral (conditional) moment generating function of the process $\{ \sqrt{v_{1,t}}Z_{1,{t}}+ \sqrt{v_{2,t}}Z_{2,t} \}$ is given by 

\begin{equation} \label{eq_prop6_prev1}
     \mathbb{E}^\mathbb{Q} \left[e^{\phi \left(\sqrt{v_{1,t}} Z_{1,t} + \sqrt{v_{2,t}} Z_{2,t} \right)} \middle | \mathcal{F}_{t-1} \right] =  \mathbb{E} \left[ e^{\phi (\sqrt{v_{1,t}} Z_{1,t} + \sqrt{v_{2,t}} Z_{2,t} )} \frac{dQ_t}{dP_t} \middle | \mathcal{F}_{t-1} \right].
\end{equation}

Using the Radon-Nikodym derivative given in equation \eqref{eq_rn_derivative} we can write the conditional expectation in \eqref{eq_prop6_prev1} as 

\begin{equation}
    \mathbb{E} \left[ e^{\phi (\sqrt{v_{1,t}} Z_{1,t} + \sqrt{v_{2,t}} Z_{2,t} )} \frac{dQ_t}{dP_t} \middle | \mathcal{F}_{t-1} \right] = \frac{\mathbb{E} \left[ e^{
    \left(\phi + \Lambda \right) (\sqrt{v_{1,t}} Z_{1,t}  \sqrt{v_{2,t}} Z_{2,t}) } \middle | \mathcal{F}_{t-1} \right]}
    {
    \mathbb{E} \left[ e^{
    \Lambda (\sqrt{v_{1,t}} Z_{1,t} + \sqrt{v_{2,t}} Z_{2,t}} ) \middle | \mathcal{F}_{t-1} \right]
    }.
\end{equation}

Since the innovations $Z_{1,t}$ and $Z_{2,t}$ are independent and based on \eqref{eq_prev3}-\eqref{eq_prev4} we finally have

\begin{align}
    \mathbb{E}^\mathbb{Q} \left[e^{\phi \left(\sqrt{v_{1,t}} Z_{1,t} +  \sqrt{v_{2,t}} Z_{2,t} \right) } \middle | \mathcal{F}_{t-1} \right]
    &= 
    \frac{\mathbb{E} \left[ e^{
    \left(\phi + \Lambda \right) ( \sqrt{v_{1,t}} Z_{1,t} + \sqrt{v_{2,t}} Z_{2,t}) } \middle | \mathcal{F}_{t-1} \right]}
    {
    \mathbb{E} \left[ e^{
    \Lambda ( \sqrt{v_{1,t}} Z_{1,t} +  \sqrt{v_{2,t}} Z_{2,t}} ) \middle | \mathcal{F}_{t-1} \right]
    } \\
    \notag
    &=
    e^{
    \frac{ {v}_{1,t} + {v}_{2,t}}{2} \left(\phi + \Lambda \right)^2 - \frac{1}{2} \Lambda^2 \left({v}_{1,t} + {v}_{2,t} 
     \right)}.
\end{align}

It can be seen that $\frac{1}{2} \phi^2 {v}_{1,t} + \phi \Lambda {v}_{1,t}$ is the exponent of the normal (conditional) moment generating function
with mean $\Lambda {v}_{1,t}$ and variance ${v}_{1,t}$, so that we denote the risk-neutral shocks as $ \Hat{Z}_{1,t} \sim N( \Lambda {v}_{1,t} , 1 )$ and  $\Hat{Z}_{2,t} \sim N( \Lambda {v}_{2,t} , 1 )$. 

Therefore, under the risk-neutral measure, the return process in \eqref{eq:1} can be written as

\begin{equation} \label{eq_risk_neutral}
    R_t = r + \lambda ( v_{1,t}  + v_{2,t} )+ \sqrt{v_{1,t}} \Hat{Z}_{1,t}+ \sqrt{v_{2,t}} \Hat{Z}_{2,t}.
\end{equation}

However, to express the normal shocks as mean-zero innovation we set

\begin{equation} \label{eq_prev_7}
      Z^*_{1,t} =  \Hat{Z}_{1,t} -  \Lambda \sqrt{v_{1,t}}, \quad
      Z^*_{2,t} =  \Hat{Z}_{2,t} - \Lambda \sqrt{v_{2,t}},
\end{equation}

\noindent so that $Z^*_{i,t} \sim N(0,1)$ under the risk-neutral measure, $i=1,2$. Then, by using \eqref{eq_Lambda} we can rewrite equation \eqref{eq_risk_neutral} as

\begin{equation*}
    R_t = r - \frac{1}{2} ( v_{1,t}  + v_{2,t}) + \sqrt{v_{1,t}} Z^*_{1,t} + \sqrt{v_{2,t}} Z^*_{2,t} .
\end{equation*}

Lastly, under the risk-neutral measure and based on equation \eqref{eq_prev_7}, the recursions in \eqref{eq:2} are written as

\begin{equation}
     \begin{bmatrix}
    v_{1,t+1} \\
    v_{2,t+1} 
    \end{bmatrix} =
    \begin{bmatrix}
        \omega_1 \\
        \omega_2
    \end{bmatrix} + 
    \begin{bmatrix}
        \beta_{11} & \beta_{12} \\
        \beta_{21} & \beta_{22} 
    \end{bmatrix} 
    \begin{bmatrix}
    v_{1,{t}} \\
    v_{2,{t}} \\
    \end{bmatrix} + 
    \begin{bmatrix}
    \alpha_{11} & 
    \alpha_{12} \\
    \alpha_{21} & \alpha_{22} \\
    \end{bmatrix}
    \begin{bmatrix}
       \left(Z^*_{1,t}-\gamma^*_1 \sqrt{v_{1,t}} \right)^2 \\
       \left(Z^*_{2,t}-\gamma^*_2 \sqrt{v_{2,t}} \right)^2 
    \end{bmatrix},
\end{equation}

\noindent where $\gamma^*_1 = \lambda + \frac{1}{2} + \gamma_1$ and $\gamma^*_2 = \lambda + \frac{1}{2} + \gamma_2$, which concludes the proof. 
$\hfill \square$

\textit{Proof of Proposition \ref{prop_option3}.} To obtain the recursive equations for $A_t (\phi)$, $B_{1,t}(\phi)$ and $B_{2,t}(\phi)$ for $t<T$, we use the tower property of the conditional expectation 

\begin{equation} \label{eq_tower}
f_t(\phi) = \mathbb{E} \left[  f_{t+1}(\phi) \middle| \mathcal{F}_t \right], 
\end{equation}

\noindent so that by using equation \eqref{eq_guess} we can express \eqref{eq_tower} as

\begin{equation} \label{eq_tower2}
    f_t(\phi) = \mathbb{E} \left[
    e^{\phi \log S_{t+1} + A_{t+1}(\phi) + B_{1,t+1}(\phi) v_{1,t+2}
    + B_{2,t+1}(\phi)v_{2,t+2} } \middle| \mathcal{F}_t \right]. 
\end{equation}

If we substitute equations \eqref{eq:1} and \eqref{eq:2} into \eqref{eq_tower2} we obtain

\begin{align*}
   f_t(\phi) = \mathbb{E} \biggl[
    &e^{\phi \left(\log S_t + r + \lambda ( v_{1,t+1} + v_{2,t+1}) + \sqrt{ v_{1,t+1} } Z_{1,t+1}  +  \sqrt{v_{2,t+1}} Z_{2,t+1} \right)  +A_{t+1}(\phi)}   \\
    & \times e^{ B_{1,t+1}(\phi) \left( \omega_1 + \beta_{11} v_{1,t+1} + \beta_{12} v_{2,t+1} + \alpha_{11} \left(Z_{1,t+1} - \gamma_{1} \sqrt{v_{1,t+1}} \right)^2 + \alpha_{12} \left(Z_{2,t+1} - \gamma_{2}\sqrt{v_{2,t+1} } \right)^2 \right) }  \\
    & \times e^{ B_{2,t+1}(\phi) \left( \omega_2 + \beta_{21} v_{1,t+1} + \beta_{22} v_{2,t+1} + \alpha_{21} \left(Z_{1,t+1} - \gamma_{1} \sqrt{v_{1,t+1}} \right)^2 + \alpha_{22} \left(Z_{2,t+1} - \gamma_{2}\sqrt{v_{2,t+1} } \right)^2 \right) } 
    \biggl| \mathcal{F}_t \biggr]. 
\end{align*}

After re-arranging terms and by completing squares and some algebra we obtain

\begin{align} \label{eq_prev_ineq_1}
   f_t(\phi) = \mathbb{E} \biggl[
    &e^{\phi (\log S_t + r ) + 
    B_{1,t+1}(\phi) \omega_1 + B_{2,t+1}(\phi)\omega_2 + A_{t+1}(\phi) + \left(\phi \lambda  + \beta_{11} + B_{1,t+1} \alpha_{11} \gamma^2_1 
    \beta_{21} + B_{2,t+1} \alpha_{21} \gamma^2_1 \right) v_{1,t+1} } 
    \\ \notag
    & \times e^{  \left(\alpha_{11} B_{1,t+1}(\phi) + \alpha_{21} B_{2,t+1}(\phi) \right) Z^2_{1,t+1}
    + \sqrt{v_{1,t+1}} 2 \gamma_1 \left( \frac{\phi}{2\gamma_1} - \alpha_{11}
    B_{1,t+1}(\phi) - \alpha_{21} B_{2,t+1}(\phi) \right) Z_{1,t+1}} 
    \\ \notag
    & \times e^{  \left(\alpha_{12} B_{1,t+1}(\phi) + \alpha_{22} B_{2,t+1}(\phi) \right) Z^2_{2,t+1}
    + \sqrt{v_{2,t+1}} 2 \gamma_2 \left( \frac{\phi}{2\gamma_2} - \alpha_{12}
    B_{1,t+1}(\phi) - \alpha_{22} B_{2,t+1}(\phi) \right) Z_{2,t+1}} 
    \\ \notag
    & \times e^{ \left(\phi \lambda  + \beta_{12} + B_{1,t+1} \alpha_{12} \gamma^2_2 
    \beta_{22} + B_{2,t+1} \alpha_{22} \gamma^2_2 \right) v_{2,t+1}}   
    \biggl| \mathcal{F}_t \biggr], 
\end{align}

\noindent so that we can make use of the following result for a standard Normal random variable $Z$:
\begin{equation} \label{eq_mgf_trick}
    \mathbb{E}\left[ e^{ aZ^2 +bZ } \right]  = e^{\frac{b^2}{2(1-2a)} - \frac{1}{2} \log(1-2a)}, \quad a < \frac{1}{2}.
\end{equation}

Using \eqref{eq_mgf_trick}, the independence between $Z_{1,t}$ and $Z_{2,t}$ and subsequently equating terms in both sides of  \eqref{eq_prev_ineq_1} we obtain:

\begin{align} \notag
    A_{t}( \phi) =&  A_{t+1}(\phi) + \phi r + B_{1,t+1}(\phi)\omega_1 + B_{2,t+1}(\phi)\omega_2 -\frac{1}{2} \log \left(1- 2 \left( B_{1,t+1}(\phi) \alpha_{11} + B_{2,t+1} (\phi) \alpha_{21} \right) \right)  \\ \notag
    &-\frac{1}{2} \log \left(1- 2 \left( B_{2,t+1}(\phi) \alpha_{22} + B_{1,t+1} (\phi)
\alpha_{12} \right) \right), 
\\[1em]
\label{eq_coeff_rico}
    B_{1,t}(\phi) =& 
    \phi \lambda + 
    \left(\beta_{11}  + \alpha_{11} \gamma_1^2 \right) B_{1,t+1}(\phi) +
     \left(\beta_{21}  + \alpha_{21} \gamma_1^2 \right)  B_{2,t+1}(\phi)  \\ \notag
     & +  
     \frac{2 \gamma^2_1 \left(
     \frac{\phi}{2\gamma_1} -
     \alpha_{11}B_{1,t+1} (\phi)
     - \alpha_{21}B_{2,t+1}(\phi)
      \right)^2}{1-2(\alpha_{11}  B_{1,t+1}(\phi) +  \alpha_{21}  B_{2,t+1}(\phi) )}, \\[1em]
      \notag
    B_{2,t}(\phi) = &
    \phi \lambda + 
    \left(\beta_{22} + \alpha_{22} \gamma^2_2 \right)  B_{2,t+1}(\phi) +
     \left( \beta_{12} + \alpha_{12} \gamma^2_2 \right)  B_{1,t+1} (\phi) \\ \notag
     +& 
     \frac{2 \gamma^2_2 \left(
     \frac{\phi}{2\gamma_2} 
     -\alpha_{22}B_{2,t+1}(\phi)
     - \alpha_{12}B_{1,t+1}(\phi)
     \right)^2}{1-2 \left(\alpha_{22}  B_{2,t+1}(\phi) +  \alpha_{12}  B_{1,t+1}(\phi) \right)}.
\end{align}

We can finally use equations \eqref{eq_coeff_rico} to calculate the coefficients recursively starting with $A_T(\phi)=B_{1,T}(\phi)=B_{2,T}(\phi)=0$.
$\hfill \square$

\bibliographystyle{unsrtnat}
\bibliography{references}  

\end{document}